\documentclass[twocol]{ametsocV6.1}
\usepackage{array}
\makeatletter
\renewenvironment{table*}{\footnotesize\@dblfloat{table}}{\end@dblfloat}

\makeatother
\runningheadauthors{Mishra et al.}
\usepackage[hidelinks]{hyperref}
\hypersetup{pdftitle={Scattering by Small Ice Crystals in Turbulence: From Orientation Statistics to Retrieval Biases},pdfauthor={Himanshu Mishra, Mohd. Meraj Khan, T. Narayana Rao, Anubhab Roy}}

\newcommand{\eps}{\varepsilon}
\newcommand{\Rey}{\mathrm{Re}_\lambda}
\newcommand{\slnphi}{\sigma^2_{\ln\phi}}
\newcommand{\pv}{\boldsymbol{p}}
\newcommand{\gh}{\hat{\boldsymbol{g}}}
\newcommand{\Sv}{\mathrm{Sv}}
\newcommand{\ahat}{\hat{\alpha}}
\newcommand{\dah}{\Delta\hat{\alpha}}
\newcommand{\PL}[1]{\langle P_{#1}\rangle}
\newcommand{\Zdr}{Z_{\mathrm{DR}}}
\newcommand{\Kdp}{K_{\mathrm{DP}}}
\newcommand{\LDR}{\mathrm{LDR}}
\newcommand{\Dmax}{D_{\max}}
\newcommand{\Xm}{X_m}

\title{Scattering by Small Ice Crystals in Turbulence: From Orientation Statistics to Retrieval Biases}

\authors{Himanshu Mishra,\aff{a} Mohd.\ Meraj Khan,\aff{a} T. Narayana Rao,\aff{b} and Anubhab Roy\aff{a}\correspondingauthor{Anubhab Roy, anubhab@iitm.ac.in}}

\affiliation{\aff{a}{Department of Applied Mechanics and Biomedical Engineering, Indian Institute of Technology Madras, Chennai, India}\\
\aff{b}{National Atmospheric Research Laboratory, Gadanki, India}}

\abstract{Ice-cloud retrievals from polarimetric radar, lidar and submillimeter radiometry commonly assume random orientation, horizontal alignment or a prescribed flutter angle. We link the orientation dynamics of ice crystals to their scattering. Fluid-inertial torque turns small settling crystals broadside on, and turbulent velocity gradients turn them away, so their orientation depends on the dissipation rate. We combine a closed-form orientation distribution for overdamped spheroids, including rotational diffusion and turbulence intermittency, with orientation-averaged Rayleigh, Rayleigh--Gans, T-matrix and specular scattering. Turbulence enters the scattering only through the Legendre moments of the orientation distribution. In the Rayleigh regime the specific differential phase follows crystal alignment, so assuming perfectly horizontal crystals biases ice-water-content retrievals low. Combining the specific differential phase with differential reflectivity limits this bias to $25\%$ for a single size of the shapes studied. For size distributions that span the alignment transition the bound can fail, because the two observables weight sizes differently. Depolarization adds sensitivity to higher-order moments and to intermittency. At finite size the second orientation moment mainly controls the scattering cross sections and the asymmetry parameter, and for optically soft crystals its contribution vanishes $54.7^\circ$ from the vertical. Specular returns from partly aligned plates should persist for lidars pointed $3^\circ$ from nadir. An exploratory analysis of EarthCARE lidar data finds a weakly depolarizing, low-lidar-ratio signature most frequent near $-11^\circ$C, where plates grow. Degree-scale tilt widths inferred from glints and lidar require larger crystals, near or beyond the validity range of the model.}

\begin{document}

\maketitle

\statement
Radar, lidar and satellite methods that measure ice clouds must assume how the crystals are oriented. The usual choices (randomly tumbling, perfectly flat, or flat with a fixed wobble) ignore the air motion. We use a recent theory for crystals settling through turbulence to show that the orientation of crystals a few tens of micrometers across depends on how turbulent the air is. We compute how this changes the radar, lidar and submillimeter signals used to estimate ice mass and crystal shape. Ice mass from the radar differential phase can be too low by an amount set by turbulence. A combined estimator limits the error for crystals of one size, but not always for mixed sizes.

\section{Introduction}

Polarimetric and multiwavelength remote sensing of ice clouds relies on assumptions about crystal orientation. Radar differential reflectivity $\Zdr$ and specific differential phase $\Kdp$ respond to the anisotropic polarizability of nonspherical crystals only when their axes are preferentially oriented. The linear depolarization ratio LDR is nonzero even for randomly oriented nonspherical crystals and depends on orientation as well as shape \citep{ryzhkov2019radar}. Specular lidar returns and glints require plates oriented within about a degree of horizontal \citep{platt1978lidar,thomas1990lidar,chepfer1999observations,breon2004horizontally,noel2010global}. They can bias cloud-phase classification and extinction retrievals, which motivated the change of CALIPSO's pointing angle from $0.3^\circ$ to $3^\circ$ off nadir \citep{hu2009calipso}. Polarization differences measured by radiometers above 100~GHz likewise require oriented particles \citep{brath2020microwave,barlakas2021introducing,mccusker2024first}. Retrievals therefore assume random orientation, horizontal alignment with random azimuth \citep{hogan2012radar,mccusker2024first}, or a Gaussian flutter of prescribed width, commonly $10^\circ$ \citep{matrosov2017atmospheric}. Reported widths span about $2^\circ$--$23^\circ$ \citep{melnikov2013axis,matrosov2021polarimetric}, and retrieved aspect ratios are apparent rather than geometric \citep{myagkov2016relationship}. Orientation also changes the radiative effect of cirrus \citep{takano1989solar,takano1989solarII,stillwell2019radiative}. Ray-tracing calculations give differences of 100--155~W~m$^{-2}$ in reflected flux between horizontally and randomly oriented crystals but assign one tilt distribution to all sizes \citep{klotzsche2006influence}. An orientation model is needed that ties the assumed state to crystal size and to the turbulence.

For crystals smaller than the Kolmogorov length and settling at small particle Reynolds number, the orientation results from a competition between alignment and randomization. Fluid inertia exerts a torque that rotates a settling spheroid toward a broadside orientation, with its largest cross section normal to gravity \citep{khayat1989inertia,dabade2015effects,roy2019inertial}, whereas turbulent velocity gradients rotate it away. Earlier models compared the velocity difference across the crystal with its fall speed \citep{klett1995orientation,breon2004horizontally}. For sub-Kolmogorov crystals the relevant scale is the Kolmogorov velocity, so the settling speed relative to it, and therefore the dissipation rate $\eps$, sets the competition \citep{gustavsson2017statistical,gustavsson2019effect,anand2020orientation,roy2023orientation}. Rotational Brownian motion adds a size-dependent randomization that sets the lower size limit of persistent orientation at tens of micrometers \citep{katz1998subsuns,kostinski2026constraining}. Observational evidence for the dependence on $\eps$ is mixed. Photographs of falling snow show broader orientation distributions in stronger near-surface turbulence \citep{garrett2015orientations}, and oriented crystals over Beijing were found mainly where $\eps<10^{-2}$~m$^2$~s$^{-3}$ \citep{wu2026observation}. Other analyses found no clear dependence \citep{westbrook2010doppler,wu2025assessment}. These studies concern crystals of 0.4--1.8~mm, an order of magnitude larger than those considered here. In the upper troposphere $\eps$ is typically $10^{-6}$--$10^{-4}$~m$^2$~s$^{-3}$, with larger values in turbulent layers \citep{karcher2025dissecting}.

Until recently the orientation density in the crossover between random and aligned states was available only from direct numerical simulations or strong-alignment asymptotics. \citet{roy2026orientation} showed that the fluid-inertial torque on a settling spheroid forms a gradient flow on the sphere of orientations. Under isotropic white forcing the stationary density is therefore a Boltzmann density. With its concentration fixed by the linear response to colored forcing, and with a lognormal superposition over dissipation rates, this closed-form density reproduces distributions measured in resolved turbulence \citep{anand2020orientation}. On the scattering side, polarimetric radar variables are commonly written through a few angular moments of the orientation distribution \citep{ryzhkov2001interpretation}. Analytical orientation averaging is established for T-matrix calculations of partially aligned particles \citep{mishchenko1991extinction}. Exact solutions exist for randomly \citep{asano1980light,mishchenko1991light} and horizontally \citep{asano1983light} oriented spheroids, and the Rayleigh--Gans approximation gives closed forms for optically soft particles \citep{vandehulst1981light,westbrook2006radar,hogan2014equation}. These treatments prescribe random, fixed, or Gaussian-flutter orientations.

Here we join the two to ask how, for a crystal of given size and habit, the quantities that polarimetric and scattering retrievals return depend on $\eps$. We take the orientation model of \citet{roy2026orientation}, add rotational Brownian motion to its closure and evaluate it in cirrus clouds. We write orientation averages through the Legendre order parameters $\PL{L}$ of the density. This implements the angular-moment framework and separates the turbulence input from the optics. The new results are the explicit dependence of Rayleigh, Rayleigh--Gans and T-matrix observables on $\eps$, the effect of size-dependent orientation on ice-water-content estimators, a face-term model of specular lidar returns, and an exploratory comparison with EarthCARE lidar data. Sections~2--6 treat the orientation state and its validity window (maximum dimensions below about 65~$\mu$m), Rayleigh polarimetry, finite size parameters, retrieval biases and specular reflection. Section~7 separates exact from closure-dependent results.

\section{Orientation distribution and alignment of settling ice crystals}
\label{sec:orient}

\begin{figure*}[tp]
\centering
\includegraphics[width=39pc]{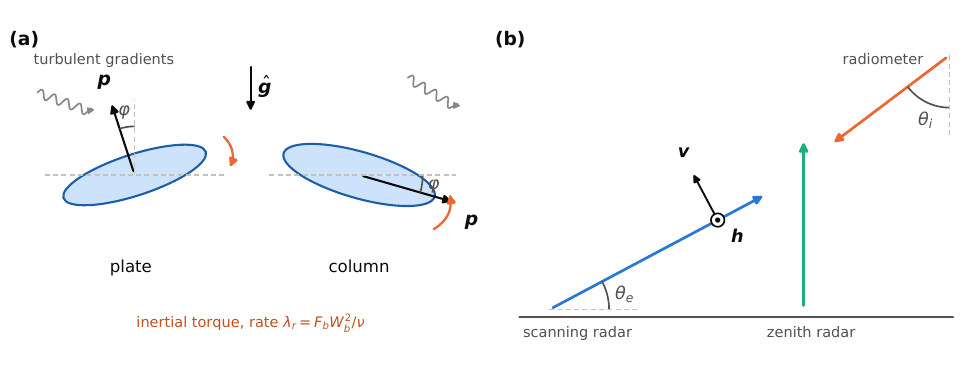}
\caption{(a) A settling plate and column tilted by $\varphi$ from their broadside attractors. The fluid-inertial torque restores the broadside orientation at rate $\lambda_r$ (orange arrows) while turbulent velocity gradients and rotational Brownian motion randomize the orientation. (b) Observation geometries: a scanning radar beam at elevation $\theta_e$ with horizontal ($\boldsymbol{h}$, out of the page) and vertical ($\boldsymbol{v}$) polarization vectors, a vertically pointing radar, and a radiometer line of sight at zenith angle $\theta_i$.}
\label{fig:schematic}
\end{figure*}

We model each ice crystal as a spheroid (density $\rho_p=917$~kg~m$^{-3}$) with symmetry semi-axis $a_\parallel$, transverse semi-axis $a_\perp$, aspect ratio $\kappa=a_\parallel/a_\perp$, and orientation $\pv$ along its symmetry axis (see figure~\ref{fig:schematic}). Its shape enters the dynamics through the Bretherton constant
\begin{equation}
\mathcal{B}=\frac{\kappa^2-1}{\kappa^2+1},
\label{eq:B}
\end{equation}
which is positive for prolate (columns) and negative for oblate (plates) crystals. We use $|\mathcal{B}|=0.75$, that is $\kappa=\sqrt{7}$ for columns and $\kappa=1/\sqrt{7}$ for plates, and characterize size by the maximum dimension $\Dmax=2a_{\max}$. The air is taken at 10~km in the standard atmosphere, $T=223$~K and $p=265$~hPa, with density $\rho_f=0.414$~kg~m$^{-3}$, dynamic viscosity $\mu_f=1.46\times10^{-5}$~Pa~s and kinematic viscosity $\nu=3.52\times10^{-5}$~m$^2$~s$^{-1}$. The Kolmogorov length, time and velocity are $\eta=(\nu^3/\eps)^{1/4}$, $\tau_K=(\nu/\eps)^{1/2}$ and $u_\eta=(\nu\eps)^{1/4}$. At $\eps=10^{-4}$~m$^2$~s$^{-3}$ they are 4.6~mm, 0.59~s and 0.77~cm~s$^{-1}$, so every crystal considered here is sub-Kolmogorov and samples a locally linear flow.

Table~\ref{tab:crystals} lists the settling properties from the anisotropic Stokes mobility (appendix~A). The volume-equivalent sphere settles at $W_{\rm eq}=2(\rho_p-\rho_f)gR_{\rm eq}^2/(9\mu_f)$, with $R_{\rm eq}=(a_\perp^2a_\parallel)^{1/3}$, and a broadside crystal at $W_b=\hat U_\perp W_{\rm eq}$ (columns, axis horizontal) or $\hat U_\parallel W_{\rm eq}$ (plates, axis vertical), with $\hat U_\parallel$ and $\hat U_\perp$ the axial and transverse settling factors. For $\Dmax\le65~\mu$m the particle Reynolds number on $W_b$ and $\Dmax$ is below 0.12, where the small-$Re_p$ inertial torque applies \citep{dabade2015effects}.

\begin{table*}[tp]
\caption{Settling and alignment properties of the crystals at 10~km ($T=223$~K, $p=265$~hPa). $W_b$: broadside fall speed; $\lambda_r$: alignment rate (\ref{eq:lamr}); $\tau_p\lambda_r$: response time $\tau_p=W_b/g$ in units of the alignment time (section~2c); $Re_p$: particle Reynolds number on $W_b$ and $\Dmax$; $\sigma_B=(D_B/\lambda_r)^{1/2}$: Brownian canting width of an aligned crystal, given where it is below $20^\circ$. Rows beyond 65~$\mu$m lie outside the overdamped window.}
\label{tab:crystals}
\begin{center}
\setlength{\tabcolsep}{4.5pt}
\begin{tabular}{rrrrrrrrrrrr}
\hline\hline
 & \multicolumn{5}{c}{plates, $\kappa=1/\sqrt7$} & & \multicolumn{5}{c}{columns, $\kappa=\sqrt7$}\\
\cline{2-6}\cline{8-12}
$\Dmax$ & $W_b$ & $\lambda_r$ & $\tau_p\lambda_r$ & $Re_p$ & $\sigma_B$ & & $W_b$ & $\lambda_r$ & $\tau_p\lambda_r$ & $Re_p$ & $\sigma_B$\\
($\mu$m) & (cm~s$^{-1}$) & (s$^{-1}$) & & & & & (cm~s$^{-1}$) & (s$^{-1}$) & & & \\
\hline
20  & 0.59 & 0.171 & $1\times10^{-4}$ & 0.003 & $17.8^\circ$ & & 0.32 & 0.044 & $1\times10^{-5}$ & 0.002 & --\\
30  & 1.32 & 0.868 & 0.001 & 0.011 & $4.3^\circ$  & & 0.73 & 0.220 & $2\times10^{-4}$ & 0.006 & $11.4^\circ$\\
40  & 2.34 & 2.74  & 0.007 & 0.027 & $1.6^\circ$  & & 1.29 & 0.697 & 0.001 & 0.015 & $4.1^\circ$\\
60  & 5.27 & 13.9  & 0.075 & 0.090 & $0.38^\circ$ & & 2.90 & 3.53  & 0.010 & 0.049 & $1.0^\circ$\\
80  & 9.37 & 43.9  & 0.42  & 0.213 & $0.14^\circ$ & & 5.16 & 11.1  & 0.059 & 0.117 & $0.37^\circ$\\
100 & 14.6 & 107   & 1.6   & 0.416 & $0.06^\circ$ & & 8.06 & 27.2  & 0.22  & 0.229 & $0.17^\circ$\\
\hline\hline
\end{tabular}
\end{center}
\end{table*}

\subsection{Orientation density}
At quasi-static slip the orientation of a settling crystal obeys \citep{roy2026orientation}
\begin{equation}
\dot{\pv}=-s_\kappa\lambda_r\,\mu\,(\mathsf{I}-\pv\pv)\cdot\gh+\boldsymbol{f}(t)+\dot{\pv}_B,
\qquad \mu=\pv\cdot\gh,
\label{eq:pdot}
\end{equation}
where $s_\kappa=\mathrm{sgn}\,\mathcal{B}$, $\gh$ is the direction of gravity (figure~\ref{fig:schematic}a), the first term is the fluid-inertial torque divided by the transverse rotational resistance, $\boldsymbol{f}$ is the Jeffery rotation by the turbulent velocity gradient \citep{jeffery1922motion} and $\dot{\pv}_B$ is rotational Brownian motion. The inertial drift is minus the surface gradient of the potential $\tfrac12 s_\kappa\lambda_r\mu^2$. It drives columns to the girdle $\mu=0$ and plates to the poles $\mu=\pm1$ at the alignment rate
\begin{equation}
\lambda_r=\frac{F_{\rm eq}W_{\rm eq}^2}{\nu}=\frac{F_bW_b^2}{\nu},
\label{eq:lamr}
\end{equation}
where $F_{\rm eq}$ and $F_b$ are dimensionless torque coefficients built from the inertial-torque amplitude of \citet{dabade2015effects} (appendix~A); $F_b=0.176$ for the plates and 0.147 for the columns. The competition between alignment and turbulence is measured by
\begin{equation}
\Lambda\equiv\lambda_r\tau_K=F_b\Sv_b^2,\qquad \Sv_b=\frac{W_b}{u_\eta},
\label{eq:Lambda}
\end{equation}
which at fixed crystal scales as $\eps^{-1/2}$ and at fixed $\eps$ as $\Dmax^4$.

If the orientation is statistically independent of the instantaneous velocity gradient, the turbulent forcing is isotropic in the tangent plane, $\langle f_if_j\rangle=\sigma_f^2(\delta_{ij}-p_ip_j)$, with
\begin{equation}
\sigma_f^2=\frac{(5+3\mathcal{B}^2)\,\eps}{60\,\nu},
\qquad
D_B=\frac{k_BT}{C_r},
\label{eq:forcing}
\end{equation}
where $D_B$ is the rotational Brownian diffusivity about a transverse axis and $C_r$ the corresponding rotational resistance. Because the inertial drift is a gradient flow and both forcings are isotropic in the tangent plane, the stationary density for white forcing is of Boltzmann form \citep{roy2026orientation}, the Watson distribution of directional statistics \citep{watson1965equatorial},
\begin{equation}
\psi(\mu)=\frac{e^{\mp b\mu^2}}{Z_\mp},\qquad
Z_-=\frac{\sqrt{\pi}\,\mathrm{erf}\sqrt{b}}{2\sqrt{b}},\qquad Z_+=\frac{e^{b}\mathcal{D}(\sqrt{b})}{\sqrt{b}},
\label{eq:boltzmann}
\end{equation}
with the upper sign for columns, the lower for plates and $\mathcal{D}$ the Dawson function. On the full sphere the density is fore--aft symmetric, $\psi(|\mu|)/2$ on $-1\le\mu\le1$; equation~(\ref{eq:boltzmann}) is its folded form, normalized on $0\le\mu\le1$. Turbulent forcing has a correlation time $\tau_c\simeq\tau_K$ (section~2c). Following \citet{roy2026orientation} we keep the Boltzmann form and fix its concentration $b$ from the exact linear response to Ornstein--Uhlenbeck forcing, which replaces the turbulent diffusivity $\sigma_f^2\tau_c$ by $\sigma_f^2\tau_c/(1+\lambda_r\tau_c)$. Brownian rotation is white and adds its diffusivity unchanged, so that
\begin{equation}
b=\frac{\lambda_r}{2}\left(\frac{\sigma_f^2\tau_c}{1+\lambda_r\tau_c}+D_B\right)^{-1}
=\frac{30\Lambda(1+\Lambda)}{5+3\mathcal{B}^2+60(1+\Lambda)D_B\tau_K}.
\label{eq:b}
\end{equation}

Without Brownian rotation this is equation~(4.6) of \citet{roy2026orientation}; the Brownian term is the extension made here. The Boltzmann form is exact for white forcing, but equation~(\ref{eq:b}) is a closure. It is exact at both ends of the range, returning the uniform density for a weak torque and the rapid-settling result of \citet{gustavsson2019effect} in the aligned limit, independently of $\tau_c$. Against stochastic simulations of the same forcing model its largest error on $\langle\mu^2\rangle$ is 15\%, near $\Lambda\approx0.45$ for a slender column (16\% with the full velocity-gradient forcing), and 8\% on $\langle1-\mu^2\rangle$ for a plate \citep{roy2026orientation}. Where the torque is weak the closure underestimates the alignment. At the weakest torques simulated, the departure from random orientation is about 1.5 times that of the closure \citep{roy2026orientation}. Near $S=1/2$ the 15\% error makes $S$ about 0.06 too low, which places $D_{1/2}$ about 4\% too high for columns, and less for plates.

The local dissipation rate fluctuates. We write it as $\phi\eps$, with $\ln\phi$ normal, $\langle\phi\rangle=1$ and variance $\slnphi$ \citep{kolmogorov1962refinement,oboukhov1962some}. Replacing $\sigma_f^2$ by $\phi\sigma_f^2$ in equation~(\ref{eq:b}) gives a local concentration $b(\phi)$. If the orientation equilibrates faster than $\phi$ changes, the density is the lognormal superposition
\begin{equation}
\begin{gathered}
\psi(\mu)=\int_0^\infty P(\phi)\,\frac{e^{\mp b(\phi)\mu^2}}{Z_\mp(b(\phi))}\,d\phi,\\
P(\phi)=\frac{1}{\phi\sqrt{2\pi\slnphi}}\exp\left[-\frac{(\ln\phi+\slnphi/2)^2}{2\slnphi}\right].
\end{gathered}
\label{eq:mixture}
\end{equation}
Without Brownian rotation $b(\phi)=b/\phi$, so in the aligned limit the superposition leaves $\langle\mu^2\rangle$ (columns) and $\langle1-\mu^2\rangle$ (plates) unchanged and multiplies the fourth moments by $e^{\slnphi}$. It reproduces the orientation distributions of oblate spheroids in direct numerical simulation over three to six decades of probability density \citep{roy2026orientation}. We take the variance from \citet{koch2002coagulation},
\begin{equation}
\begin{gathered}
\slnphi=-0.15+0.25\ln\left(3.1\Rey^{1/2}+0.1\Rey^{3/2}\right),\\
\Rey=15^{1/2}\frac{(\eps L)^{2/3}}{(\nu\eps)^{1/2}},
\end{gathered}
\label{eq:sigma2}
\end{equation}
with an outer scale $L=20$~m. In stably stratified air the Ozmidov scale $(\eps/N^3)^{1/2}$ can limit $L$ \citep{riley2008stratified}. For $N=0.012$~s$^{-1}$ it lowers $e^{\slnphi}$ from 5.0 to 2.4 at $\eps=10^{-6}$ and from 6.6 to 5.2 at $10^{-4}$~m$^2$~s$^{-3}$, so the superposition overstates intermittency in weak turbulence. Near $10^{-7}$~m$^2$~s$^{-3}$ the buoyancy Reynolds number $\eps/(\nu N^2)$ is about 20, and isotropic forcing becomes questionable. The superposition requires the dissipation sampled by a crystal to vary slowly compared with $1/\lambda_r$. This holds for the large-scale part of the log-dissipation variance, which varies on the integral time. Its Kolmogorov-scale part instead acts as additional fast noise when $\Lambda<1$. We therefore use two limiting scenarios, uniform dissipation (W; equations~(\ref{eq:boltzmann}) and (\ref{eq:b})) as the reference and the lognormal superposition (MA; equation~(\ref{eq:mixture})). Their difference measures the sensitivity to intermittency and is not a bound. The other closure of \citet{roy2026orientation}, with the correlation time following the local Kolmogorov time, gives values between them, or within 0.01 in $S$, in the crossover.

\subsection{Alignment parameter, canting width and two scaling laws}

Every orientation-averaged quantity in sections~3 and~4 depends on the density through its order parameters $\PL{L}$, the averages of the Legendre polynomials $P_L(\mu)$ over the full sphere. The odd ones vanish by fore--aft symmetry, and the even ones equal $\int_0^1\psi(\mu)P_L(\mu)\,d\mu$ with the folded density. We measure alignment by
\begin{equation}
S=\frac{\PL{2}}{P_2(\mu_\ast)}=
\begin{cases}\PL{2}, & \text{plates } (\mu_\ast=1),\\ -2\PL{2}, & \text{columns } (\mu_\ast=0),\end{cases}
\label{eq:S}
\end{equation}
which is one for broadside crystals and zero for random ones. We also define an equivalent canting width, as used in retrieval studies, through $\sigma_{\rm eq}^2=\langle\varphi^2\rangle/2$ for plates and $\sigma_{\rm eq}^2=\langle\varphi^2\rangle$ for columns, where $\varphi$ is the angle between $\pv$ and its attractor. The factors differ because a plate has two restoring tilt directions about its vertical attractor, whereas a column has one, out of the horizontal plane, and an attractor that is degenerate in azimuth. In the aligned limit $\sigma_{\rm eq}$ is the standard deviation of the Gaussian canting distributions used in retrievals, $\sigma_{\rm eq}^2=1/(2b)$ for both shapes.

\begin{figure*}[tp]
\centering
\includegraphics[width=1.0\textwidth]{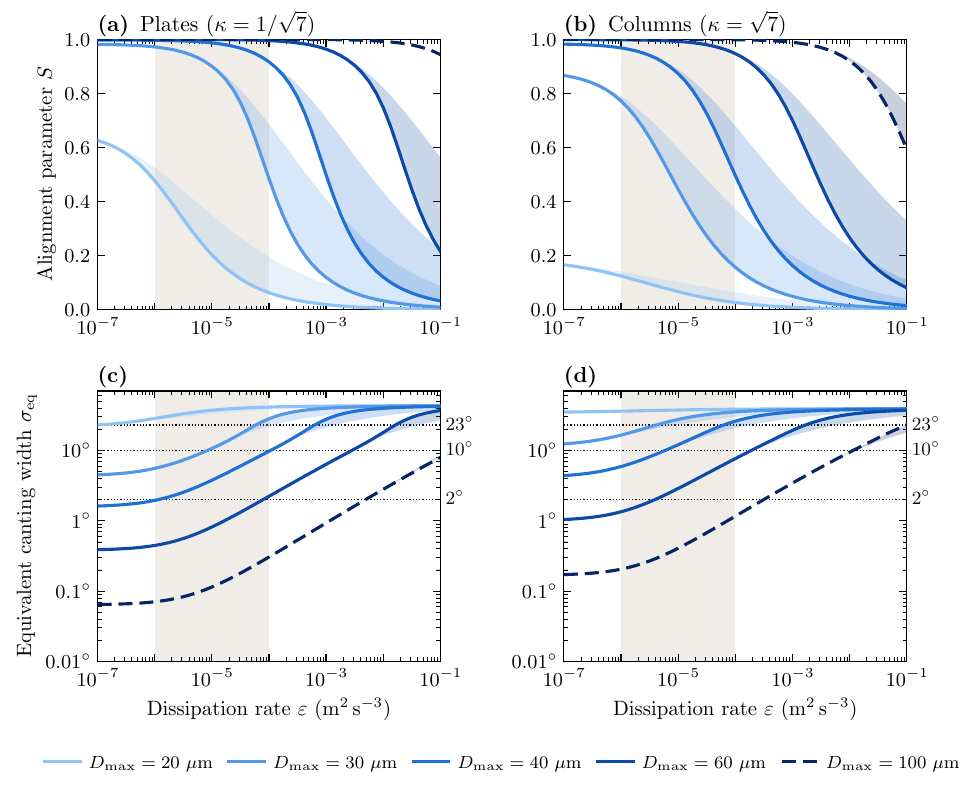}
\caption{Alignment of plates (left) and columns (right) of $|\mathcal{B}|=0.75$ against the mean dissipation rate, in air at 10~km. (a,b) Alignment parameter $S$ from equation~(\ref{eq:S}); (c,d) equivalent canting width, with the $10^\circ$ flutter commonly assumed in retrievals and the $2^\circ$ and $23^\circ$ limits of reported values marked by horizontal lines. Solid lines: uniform dissipation (W) including Brownian rotation; shading extends to the lognormal superposition (MA). The dashed curve ($\Dmax=100~\mu$m) lies outside the overdamped window. The vertical band marks $\eps=10^{-6}$--$10^{-4}$~m$^2$~s$^{-3}$, typical of the upper troposphere.}
\label{fig:alignment}
\end{figure*}

Figure~\ref{fig:alignment} shows $S$ and $\sigma_{\rm eq}$ against $\eps$ for $\Dmax=20$--100~$\mu$m. The transition is sharp in size: at $\eps=10^{-4}$~m$^2$~s$^{-3}$, plates of 30, 40 and 60~$\mu$m have $S=0.49$, 0.92 and 0.995 and canting widths of $28^\circ$, $9.7^\circ$ and $2.2^\circ$. Intermittency matters mainly within this crossover, where quiet regions of the flow allow partial alignment; for the 30-$\mu$m plate it raises $S$ to 0.69.

Two scaling laws summarize the figure. In the aligned regime $b\gg1$, equation~(\ref{eq:b}) gives
\begin{equation}
\begin{gathered}
\sigma_{\rm eq}^2=\frac{5+3\mathcal{B}^2}{60\,\Lambda(1+\Lambda)}+\frac{D_B}{\lambda_r},
\quad\text{so that}\\
\sigma_{\rm eq}\simeq C_\sigma\left(\frac{u_\eta}{W_b}\right)^2,\quad C_\sigma=\frac{1}{F_b}\left(\frac{5+3\mathcal{B}^2}{60}\right)^{1/2},
\end{gathered}
\label{eq:sigeq}
\end{equation}
when $\Lambda\gg1$ and Brownian rotation is negligible. $C_\sigma=1.90$~rad for the plates and 2.27~rad for the columns, so a plate falling at ten Kolmogorov velocities flutters by about $1.1^\circ$. The width grows as $\eps^{1/2}$ and, at fixed shape, falls as $\Dmax^{-4}$. Brownian rotation contributes $D_B/\lambda_r\propto \Dmax^{-7}$ and dominates only for the smallest aligned crystals in weak turbulence. For a 40-$\mu$m plate the Brownian width alone is $1.6^\circ$ (table~\ref{tab:crystals}), larger than the turbulent width for $\eps\lesssim2\times10^{-6}$~m$^2$~s$^{-3}$.

The second law locates the transition. With uniform dissipation and no Brownian rotation, $S=1/2$ at $\Lambda_{1/2}=0.513$ for the plates and 0.408 for the columns, the broadside settling number is
\begin{equation}
\Sv_{b,1/2}=\left(\frac{\Lambda_{1/2}}{F_b}\right)^{1/2}=1.71\ \text{(plates)},\quad 1.66\ \text{(columns)}.
\label{eq:half}
\end{equation}
A crystal is half aligned when it falls at about 1.7 Kolmogorov velocities. Because $W_b\propto(\rho_p-\rho_f)g\Dmax^2/\mu_f$ at fixed shape, the half-alignment size scales as
\begin{equation}
D_{1/2}\propto\left[\frac{\mu_f}{(\rho_p-\rho_f)g}\right]^{1/2}(\nu\eps)^{1/8}.
\label{eq:Dhalf}
\end{equation}
 Table~\ref{tab:half} lists $D_{1/2}$. Without Brownian rotation it runs from 17 to 71~$\mu$m for plates and from 22 to 95~$\mu$m for columns over $\eps=10^{-6}$--$10^{-1}$~m$^2$~s$^{-3}$.

\begin{table*}[tp]
\caption{Half-alignment size $D_{1/2}$ ($\mu$m), at which $S=1/2$, for uniform dissipation without and with rotational Brownian motion, and for the lognormal superposition from equation~(\ref{eq:mixture}) including Brownian motion.}
\label{tab:half}
\begin{center}
\begin{tabular}{lrrrrrrr}
\hline\hline
$\eps$ (m$^2$~s$^{-3}$) & $10^{-7}$ & $10^{-6}$ & $10^{-5}$ & $10^{-4}$ & $10^{-3}$ & $10^{-2}$ & $10^{-1}$\\
\hline
plates, no Brownian & 12.6 & 16.9 & 22.5 & 30.0 & 40.0 & 53.3 & 71.1\\
plates, Brownian & 19.1 & 20.2 & 23.5 & 30.2 & 40.0 & 53.3 & 71.1\\
plates, intermittent & 19.1 & 19.8 & 21.6 & 25.7 & 32.8 & 42.8 & 56.1\\
columns, no Brownian & 16.8 & 22.4 & 29.9 & 39.9 & 53.2 & 70.9 & 94.5\\
columns, Brownian & 24.3 & 26.0 & 30.9 & 40.1 & 53.2 & 70.9 & 94.5\\
columns, intermittent & 24.2 & 25.4 & 28.1 & 33.7 & 43.2 & 56.4 & 74.0\\
\hline\hline
\end{tabular}
\end{center}
\end{table*}

\subsection{Regime map and validity window}

\begin{figure*}[tp]
\centering
\includegraphics[width=1\linewidth]{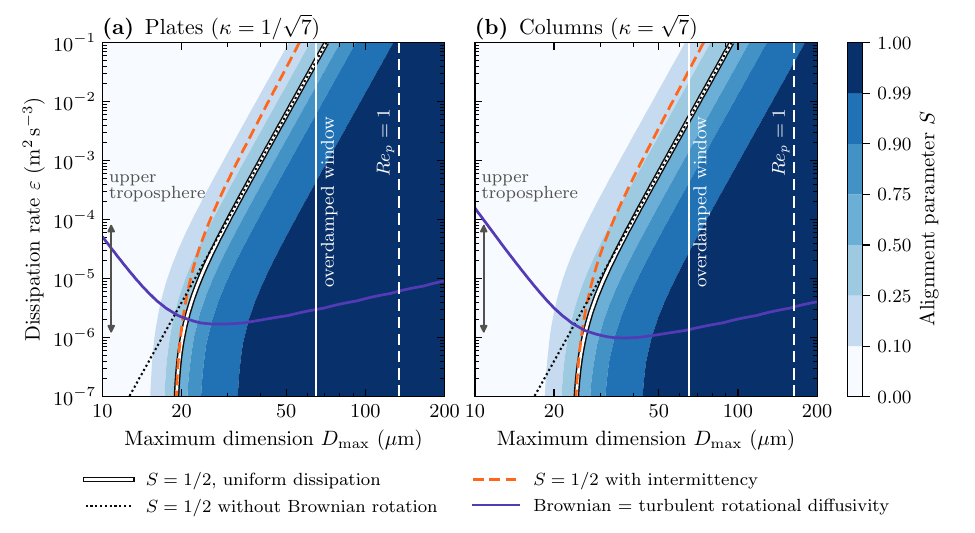}
\caption{Alignment parameter $S$ over maximum dimension and dissipation rate for (a) plates and (b) columns. White line: $S=1/2$ for uniform dissipation with Brownian rotation; dotted line: the same without Brownian rotation, following $D_{1/2}\propto\eps^{1/8}$; dashed orange line: $S=1/2$ with the lognormal superposition; purple line: equal Brownian and turbulent rotational diffusivities, Brownian rotation dominating below it. Vertical white lines mark the edge of the overdamped window, $\Dmax=65~\mu$m (section~2c), and $Re_p=1$; the arrow marks the upper-tropospheric range of $\eps$.}
\label{fig:map}
\end{figure*}

Figure~\ref{fig:map} maps $S$ in the $(\Dmax,\eps)$ plane: turbulence randomizes small crystals in strong turbulence and Brownian rotation in weak turbulence, and larger crystals are aligned.

Equation~(\ref{eq:pdot}) neglects particle inertia. The overdamped description requires the particle response time, $\tau_p\simeq W_b/g$ for broadside settling, to be short compared with the alignment time $1/\lambda_r$. It must also be short compared with the time $\ell\eta/W_b$ to settle through a correlation length of the velocity gradients, with $\ell\approx10$ or more \citep{gustavsson2021effect}. When $\tau_p$ is finite the slip velocity lags the flow, its direction tilts by about $\tau_p$ times the velocity gradient, and the crystal aligns with the tilted slip. In a linear model of this mechanism with isotropic gradient statistics (Supplement S1) the tilt variance changes by the factor $1-0.82x+1.20x^2$, with $x=\tau_p\lambda_r$. The quadratic term is the ratio of the asymptotic variances of \citet[their eqs.~24a,b]{gustavsson2021effect}. The linear term arises because the slip tilt and the Jeffery rotation oppose each other. At 10~km $x\propto\Dmax^6$ reaches 0.12 for plates and 0.017 for columns at $\Dmax=65~\mu$m (Supplement S1; table~\ref{tab:crystals} gives other sizes), where inertia lowers the tilt variance by 8\% and 1.4\%. The factor returns to one at $x\approx0.69$ (87 and 121~$\mu$m) and grows rapidly beyond $x\approx1$ (92 and 128~$\mu$m). For 65-$\mu$m crystals the Stokes number $\tau_p/\tau_K$ and the ratio $\tau_pW_b/(\ell\eta)$ stay below 0.11 for $\eps\le10^{-2}$~m$^2$~s$^{-3}$ (Supplement S1). We take $\Dmax\le65~\mu$m as the overdamped window, set by the plates and conservative for columns. It contains the half-alignment size for $\eps\lesssim5\times10^{-2}$ (plates) and $5\times10^{-3}$~m$^2$~s$^{-3}$ (columns). The same time scales justify $\tau_c\simeq\tau_K$ in the crossover (Supplement S1). For crystals beyond $x\approx1$ the overdamped model overestimates alignment: in finite-Stokes-number simulations the rms tilt of columns with semi-major axes of 100--200~$\mu$m exceeds its prediction by factors of 3--16 \citep{pumir2026statistical}.

\section{Rayleigh polarimetric operators}
\label{sec:rayleigh}

\subsection{Reduction to two order parameters}

At centimeter and millimeter wavelengths the crystals of section~2 are Rayleigh scatterers: the size parameter on the maximum dimension, $\Xm=ka_{\max}$ with $k=2\pi/\lambda$, is below 0.07 at 94~GHz for $\Dmax\le65~\mu$m. A crystal scatters as a dipole with polarizability volume $(V/4\pi)\,\hat{\boldsymbol{\alpha}}$,
\begin{equation}
\begin{gathered}
\hat{\boldsymbol{\alpha}}=\ahat_\perp\mathsf{I}+\dah\,\pv\pv,\\
\ahat_j=\frac{\epsilon_r-1}{1+L_j(\epsilon_r-1)},\qquad \dah=\ahat_\parallel-\ahat_\perp,
\end{gathered}
\label{eq:alpha}
\end{equation}
with $L_\parallel$ and $L_\perp=(1-L_\parallel)/2$ the depolarization factors of the spheroid \citep{bohren2008absorption} and $\epsilon_r$ the permittivity of ice. For $\epsilon_r=3.14+0.004i$ \citep[$T=223$~K, 94~GHz;][]{matzler2006microwave} the normalized polarizabilities are $(\ahat_\parallel,\ahat_\perp)=(0.935,1.504)$ for the plates and $(1.686,1.107)$ for the columns.

For a beam at elevation $\theta_e$ (figure~\ref{fig:schematic}b), with horizontal and vertical polarization vectors $\boldsymbol{h}$ and $\boldsymbol{v}$, the Rayleigh amplitudes depend on $\boldsymbol{e}\cdot\boldsymbol{\hat{\alpha}}\cdot\boldsymbol{e}$ for polarization directions $\boldsymbol{e}$, and hence on moments such as $\langle(\pv\cdot\boldsymbol{e})^2\rangle$ and $\langle(\pv\cdot\boldsymbol{e})^4\rangle$, as is standard in radar polarimetry \citep{ryzhkov2001interpretation}. For densities axisymmetric about the vertical the addition theorem for Legendre polynomials gives
\begin{equation}
\begin{aligned}
\langle(\pv\cdot\boldsymbol{e})^2\rangle&=\dfrac13+\dfrac{2}{3}\PL{2}P_2(e_z),\\
\langle(\pv\cdot\boldsymbol{e})^4\rangle&=\dfrac15+\dfrac47\PL{2}P_2(e_z)+\dfrac{8}{35}\PL{4}P_4(e_z),
\end{aligned}
\label{eq:addition}
\end{equation}
together with the mixed average $\langle(\pv\cdot\boldsymbol{h})^2(\pv\cdot\boldsymbol{v})^2\rangle$ of appendix~B. Turbulence therefore enters Rayleigh polarimetry through $\PL{2}$ and $\PL{4}$ alone. The reflectivities are
\begin{equation}
\begin{aligned}
Z_{hh}&\propto|\ahat_\perp|^2+2\,\mathrm{Re}(\ahat_\perp^\ast\dah)\langle(\pv\cdot\boldsymbol{h})^2\rangle\\
&\qquad+|\dah|^2\langle(\pv\cdot\boldsymbol{h})^4\rangle,\\
Z_{hv}&\propto|\dah|^2\langle(\pv\cdot\boldsymbol{h})^2(\pv\cdot\boldsymbol{v})^2\rangle,
\end{aligned}
\label{eq:Z}
\end{equation}
with $Z_{vv}$ obtained by $\boldsymbol{h}\to\boldsymbol{v}$, the linear ratio $Z_{\rm dr}=Z_{hh}/Z_{vv}$, its value in decibels $\Zdr=10\log_{10}Z_{\rm dr}$, and $\LDR=Z_{hv}/Z_{hh}$. Quantities linear in the amplitude involve $\PL{2}$ alone. The specific differential phase,
\begin{equation}
\begin{gathered}
\Kdp=\frac{180}{\lambda}\,\frac{\mathrm{IWC}}{\rho_p}\,\mathrm{Re}(-\dah)\cos^2\theta_e\,\PL{2}\\
(\mathrm{deg~per~unit~length}),
\end{gathered}
\label{eq:kdp}
\end{equation}
is exactly proportional to $\PL{2}$, so that $\Kdp/\Kdp^{\rm BO}=S$ for any orientation density, BO denoting broadside alignment. The same holds for extinction along a slanted path. For a line of sight at zenith angle $\theta_i$ both the absorption, $\propto\mathrm{Im}\,\langle\boldsymbol{e}\cdot\hat{\boldsymbol{\alpha}}\cdot\boldsymbol{e}\rangle$, and the scattering, $\propto\langle|\hat{\boldsymbol{\alpha}}\cdot\boldsymbol{e}|^2\rangle$, are linear in $\langle(\pv\cdot\boldsymbol{e})^2\rangle$, so the difference between vertically and horizontally polarized extinction is proportional to $\PL{2}\sin^2\theta_i$. Normalized by the mean extinction, which depends only weakly on $\PL{2}$, it gives the dichroism $(\sigma_V-\sigma_H)/\bar\sigma$ of section~4. These relations are exact within the Rayleigh approximation for any density axisymmetric about the vertical and do not require the Boltzmann form. The closures of section~2 supply $\PL{2}$ and $\PL{4}$ as functions of $\eps$.

\subsection{Dependence on dissipation rate}

\begin{figure*}[tp]
\centering
\includegraphics[width=1\linewidth]{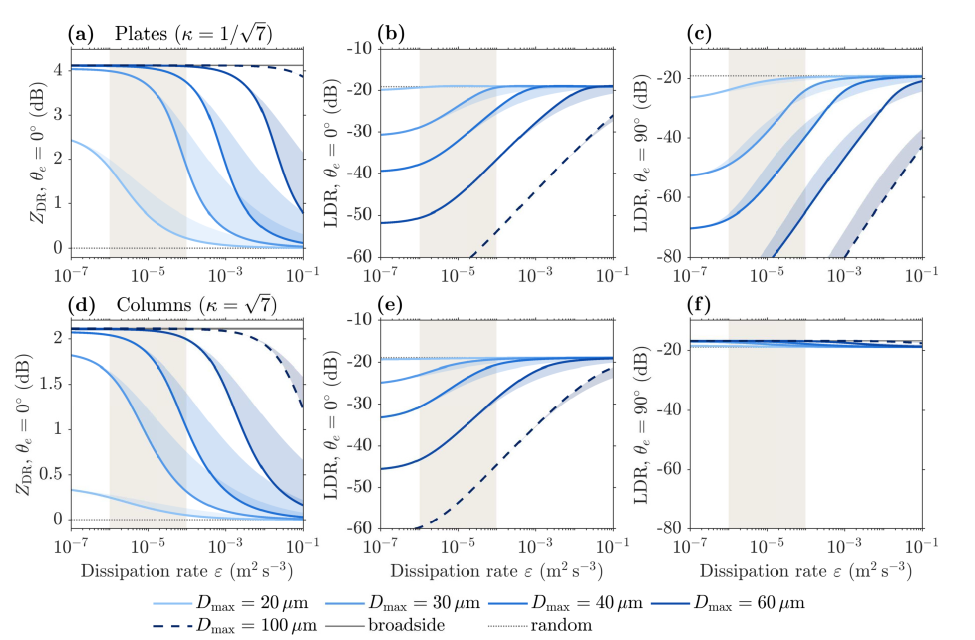}
\caption{Rayleigh polarimetric variables against dissipation rate for plates (top) and columns (bottom): (a,d) $\Zdr$ at horizontal incidence; (b,e) LDR at horizontal incidence; (c,f) LDR at vertical incidence. Solid gray lines represent broadside alignment and dotted lines random orientation. Lines and shading as in figure~\ref{fig:alignment}. Ice permittivity $\epsilon_r=3.14+0.004i$.}
\label{fig:polar}
\end{figure*}

Figure~\ref{fig:polar} shows $\Zdr$ at horizontal incidence and LDR at horizontal and vertical incidence. Broadside plates have $Z_{\rm dr}=|\ahat_\perp/\ahat_\parallel|^2$, or $\Zdr=4.13$~dB, and broadside columns, which remain random in azimuth, 2.11~dB; both fall toward 0~dB with $S$.

The depolarization ratio behaves differently. For aligned plates at horizontal incidence the cross-polar return comes from the tilt alone, $\langle(\pv\cdot\boldsymbol{h})^2(\pv\cdot\boldsymbol{v})^2\rangle=\tfrac12\langle\mu^2(1-\mu^2)\rangle\simeq\sigma_{\rm eq}^2$, so that
\begin{equation}
\begin{aligned}
\LDR(0^\circ)&\simeq\left|\frac{\dah}{\ahat_\perp}\right|^2\sigma_{\rm eq}^2,\qquad
\LDR(90^\circ)\simeq\left|\frac{\dah}{\ahat_\perp}\right|^2 \Phi\,\sigma_{\rm eq}^4,\\
\Phi&=\frac{A^2e^{\slnphi}+2AC+C^2}{(A+C)^2}\qquad\text{(aligned plates)},
\end{aligned}
\label{eq:ldr}
\end{equation}
with $|\dah/\ahat_\perp|^2=0.143$ for the plates. Here $A$ and $C$ are the turbulent and Brownian parts of the local inverse concentration, $1/b(\phi)=A\phi+C$ (appendix~B), so that $\Phi=1$ for uniform dissipation and $1\le\Phi\le e^{\slnphi}$ for the superposition. These are the leading terms of an expansion about the attractor that requires $\sigma_{\rm eq}\lesssim3^\circ$ (Supplement); figure~\ref{fig:polar} uses the full densities. The horizontal-incidence LDR therefore measures the canting width, rising from $-50$~dB for 60-$\mu$m plates at $\eps=10^{-6}$~m$^2$~s$^{-3}$ toward the random-orientation value of $-19$~dB. It exceeds a cross-polar isolation of $-30$~dB only for $\sigma_{\rm eq}\gtrsim5^\circ$. At vertical incidence the cross-polar return of a plate needs two powers of the tilt, so $\LDR(90^\circ)$ scales as $\sigma_{\rm eq}^4$ and depends on intermittency through the fourth moment $\langle(1-\mu^2)^2\rangle$. The ratio $\LDR(90^\circ)/\LDR(0^\circ)^2=\Phi|\ahat_\perp/\dah|^2$ is a polarimetric measure of the flatness of the tilt distribution \citep{roy2026orientation}. For 60-$\mu$m plates at $\eps=10^{-4}$~m$^2$~s$^{-3}$ $\Phi$ is within 5\% of $e^{\slnphi}$, but at $10^{-6}$~m$^2$~s$^{-3}$ it is only 1.3--1.5 for 40- and 60-$\mu$m plates, because Brownian rotation supplies most of the tilt. For plates with $S>0.9$, $\LDR(90^\circ)$ stays below about $-38$~dB ($-33.5$~dB with intermittency), near or below the cross-polar isolation of cloud radars \citep[see also][]{wu2025assessment}. For columns the vertical-incidence LDR stays between $-19$ and $-17$~dB for all $\eps$ (figure~\ref{fig:polar}f), because a horizontal column is random in azimuth.

\section{Finite size parameter}
\label{sec:tmatrix}

At wavelengths below about 0.4~mm the larger crystals of section~2 reach size parameters of order one ($\Xm=1$ for $\Dmax=65~\mu$m at 0.2~mm). Internal interference then adds an orientation dependence through the projection of the crystal on the scattering vector. Results at a fixed $\Xm$ for crystals of different size refer to different wavelengths, $\lambda=\pi\Dmax/\Xm$.

\subsection{Rayleigh--Gans theory}

For an optically soft crystal, $|m-1|\ll1$ and $2\Xm|m-1|\ll1$, the Rayleigh--Gans approximation treats every volume element as an independent dipole \citep{vandehulst1981light}. The amplitude scattering matrix is diagonal, with $S_1=ik^3(m-1)V\mathcal{F}(U)/2\pi$ and $S_2=S_1\cos\vartheta$ at scattering angle $\vartheta$, where for a spheroid \citep{buehler1990light}
\begin{equation}
\begin{gathered}
\mathcal{F}(U)=\frac{3(\sin U-U\cos U)}{U^3},\\
U=2X\sin\dfrac{\vartheta}{2}\left(1+r\mu_q^2\right)^{1/2},\\
X=ka_\perp,\qquad r=\kappa^2-1.
\end{gathered}
\label{eq:formfactor}
\end{equation}
Here $\mathcal{R}=\mathcal{F}^2$ is the form factor and $\mu_q=\pv\cdot\hat{\boldsymbol{q}}$ the projection of the crystal axis on the unit scattering vector $\hat{\boldsymbol{q}}\propto\hat{\boldsymbol{k}}_s-\hat{\boldsymbol{k}}_i$. Used for ice aggregates at radar wavelengths \citep{westbrook2006radar,tyynela2013applicability,hogan2014equation,leinonen2018evaluation}, it is not accurate for solid ice, for which $2\Xm|m-1|=1.6$ at $\Xm=1$, but it exposes how orientation enters at finite size. Expanding $\mathcal{R}=\sum_{n} R_nP_n(\mu_q)$ and applying the addition theorem gives
\begin{equation}
\langle\mathcal{R}\rangle=\sum_{n\ {\rm even}}\PL{n}\,R_n\!\left(X\sin\dfrac{\vartheta}{2},r\right)P_n(\gh\cdot\hat{\boldsymbol{q}}).
\label{eq:rholeg}
\end{equation}
Turbulence enters only through the order parameters, the size and shape of the crystal only through the coefficients $R_n$, and the geometry only through the angle between the scattering vector and the vertical. For unpolarized incidence the scattering and backscatter cross sections, normalized by their small-particle values, and the asymmetry parameter are $\sigma_s/\sigma_0=(3/16\pi)\int\langle\mathcal{R}\rangle(1+\cos^2\vartheta)\,d\Omega$, $\sigma_b/\sigma_{b0}=\langle\mathcal{R}\rangle$ at $\vartheta=\pi$ and $g=(3/16\pi)\int\langle\mathcal{R}\rangle(1+\cos^2\vartheta)\cos\vartheta\,d\Omega/(\sigma_s/\sigma_0)$.

Expanding $\mathcal{R}=1-U^2/5+O(U^4)$ and using $\langle\mu_q^2\rangle=\tfrac13+\tfrac23\PL{2}P_2(\gh\cdot\hat{\boldsymbol{q}})$ gives, with $\theta_i$ the zenith angle of the beam,
\begin{equation}
\begin{aligned}
\frac{\sigma_s}{\sigma_0}&=1-\frac{2X^2}{5}\left(1+\frac{r}{3}\right)-\frac{11}{75}rX^2\PL{2}P_2(\cos\theta_i),\\
g&=\frac{4X^2}{25}\left(1+\frac{r}{3}\right)+\frac{8}{75}rX^2\PL{2}P_2(\cos\theta_i),\\
\frac{\sigma_b}{\sigma_{b0}}&=1-\frac{4X^2}{5}\left[1+\frac{r}{3}+\frac{2r}{3}\PL{2}P_2(\cos\theta_i)\right],
\end{aligned}
\label{eq:rgaX2}
\end{equation}
with errors of $O(X^4)$ (appendix~C); for columns $X=\Xm/\kappa$. At this order orientation enters only through $\PL{2}$, so the three quantities are linear in $S$, as $\Kdp$ is in the Rayleigh limit. The orientation term is proportional to $rX^2=k^2(a_\parallel^2-a_\perp^2)$. It carries the factor $P_2(\cos\theta_i)$, which changes sign between vertical and horizontal incidence and vanishes at $\theta_i=54.7^\circ$. The last property holds for any quantity symmetric about the beam that depends on orientation only through $\langle\pv\pv\rangle$, such as a cross section, the asymmetry parameter, or the backscatter for unpolarized incidence. Such a quantity depends on $\langle(\pv\cdot\hat{\boldsymbol{k}}_i)^2\rangle=\tfrac13+\tfrac23\PL{2}P_2(\cos\theta_i)$, which is the same for every orientation density at that angle.

\begin{figure*}[tp]
\centering
\includegraphics[width=1\linewidth]{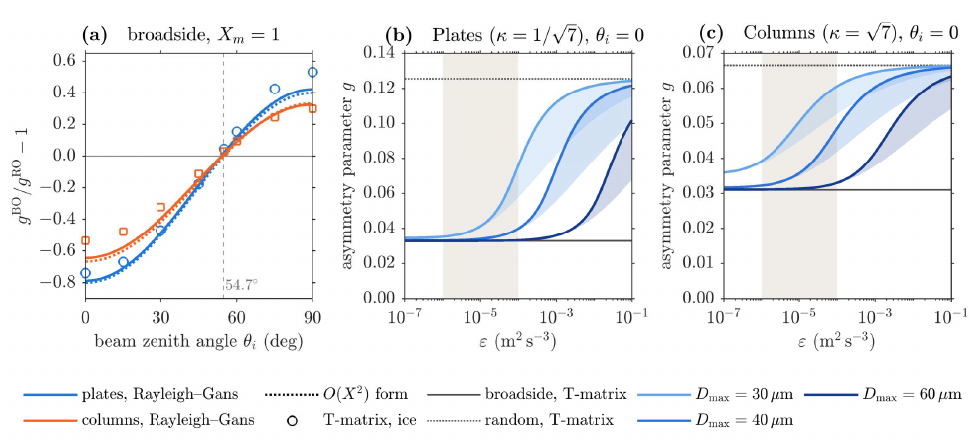}
\caption{Asymmetry parameter at $\Xm=1$. (a) Relative difference between broadside and random orientation, $g^{\rm BO}/g^{\rm RO}-1$, against the zenith angle $\theta_i$ of the beam for plates (blue, circles) and columns (orange, squares). Solid: exact Rayleigh--Gans average (equation~(\ref{eq:rholeg})); dotted: its $O(X^2)$ form (equation~(\ref{eq:rgaX2})), both in the limit $m\to1$; open symbols: T-matrix for ice with $m=1.79+0.01i$. Each curve is normalized by its own random-orientation value. The dashed vertical line marks $54.7^\circ$. (b,c) T-matrix asymmetry parameter at vertical incidence against dissipation rate for plates and columns; solid and dotted gray lines: broadside and random orientation. Lines and shading as in figure~\ref{fig:alignment}.}
\label{fig:asym}
\end{figure*}

For solid ice the property is approximate (figure~\ref{fig:asym}a). At $\Xm=1$ the broadside asymmetry parameter crosses its random-orientation value at $\theta_i=52.9^\circ$ (plates) and $52.8^\circ$ (columns) in the T-matrix, against $53.9^\circ$ and $54.1^\circ$ for the exact Rayleigh--Gans average. As $\Xm\to0$ the T-matrix crossing tends to about $53.8^\circ$ and $53.6^\circ$. The anisotropic internal field of solid ice makes $\sigma_s$ orientation dependent already as $X\to0$ and brings higher order parameters into the $O(X^2)$ terms. At $\Xm=1$ and vertical incidence broadside plates have $g=0.033$ against 0.126 for random orientation, and broadside columns 0.031 against 0.067 (T-matrix; figure~\ref{fig:asym}b,c). Turbulence moves $g$ toward its random value along the alignment curve. For beams more than about $53^\circ$ from the vertical the sign reverses, and aligned crystals scatter more strongly forward than random ones.

\subsection{Exact orientation averaging of T-matrix kernels}

We compute the scattering by solid ice with the extended-boundary-condition T-matrix method for spheroids \citep{mishchenko1991light,mishchenko1998capabilities}, through the interface of \citet{leinonen2014high}, for $m=1.79+0.01i$, the value of \citet{warren2008optical} near 0.7~mm at 266~K. With the more absorbing indices at 0.1--0.2~mm, where crystals inside the window reach $\Xm=1$--2, the departures from $S$ discussed in section~4c change by at most 0.004 at $\Xm=1$ (Supplement). For a beam along the vertical, or for any fixed incidence after averaging over the crystal azimuth, each scattering kernel $F(\mu)$ of a T-matrix truncated at order $N$ is a smooth even function of $\mu$. Its Legendre coefficients fall below $10^{-14}$ of the largest beyond degree $4N$. The orientation average is then exact to that accuracy on the $M_h=2N+1$ positive nodes $\mu_k$ of the $(4N+2)$-point Gauss--Legendre rule on $[-1,1]$,
\begin{equation}
\begin{gathered}
\langle F\rangle=\sum_{k=1}^{M_h}W_kF(\mu_k),\\
W_k=\varpi_k\sum_{L\ {\rm even}}^{4N}(2L+1)\PL{L}P_L(\mu_k),
\end{gathered}
\label{eq:weights}
\end{equation}
with $\varpi_k$ the corresponding Gauss weights; the rule is exact to degree $8N+3$, which covers the product of the truncated kernel and density. As in equation~(\ref{eq:rholeg}), the weights carry the entire turbulence input. Kernels computed once for a given shape, refractive index and size parameter therefore serve every dissipation rate and intermittency level, and every crystal size at that size parameter. Equation~(\ref{eq:weights}) agrees with direct quadrature on 400 nodes graded toward the attractor to better than $10^{-13}$. This and the checks of the T-matrix code against Mie theory and the Rayleigh--Gans limit (appendix~D) verify the numerical evaluation, not the orientation closure. The observables are the extinction and scattering cross sections, the asymmetry parameter, the backscatter cross section and LDR for a vertical beam, and the extinction for vertical and horizontal polarization along a line of sight at $\theta_i=53^\circ$. We express the backscatter through the ratio $\rho_b=\sigma_b/\sigma_b^{\rm R}$ to the Rayleigh backscatter of the same orientation density, so that the dual-wavelength ratio between a Rayleigh channel and a channel at $\Xm$ is $\mathrm{DWR}=-10\log_{10}\rho_b$.

\subsection{Which observables follow the alignment parameter}

\begin{figure*}[tp]
\centering
\includegraphics[width=1\linewidth]{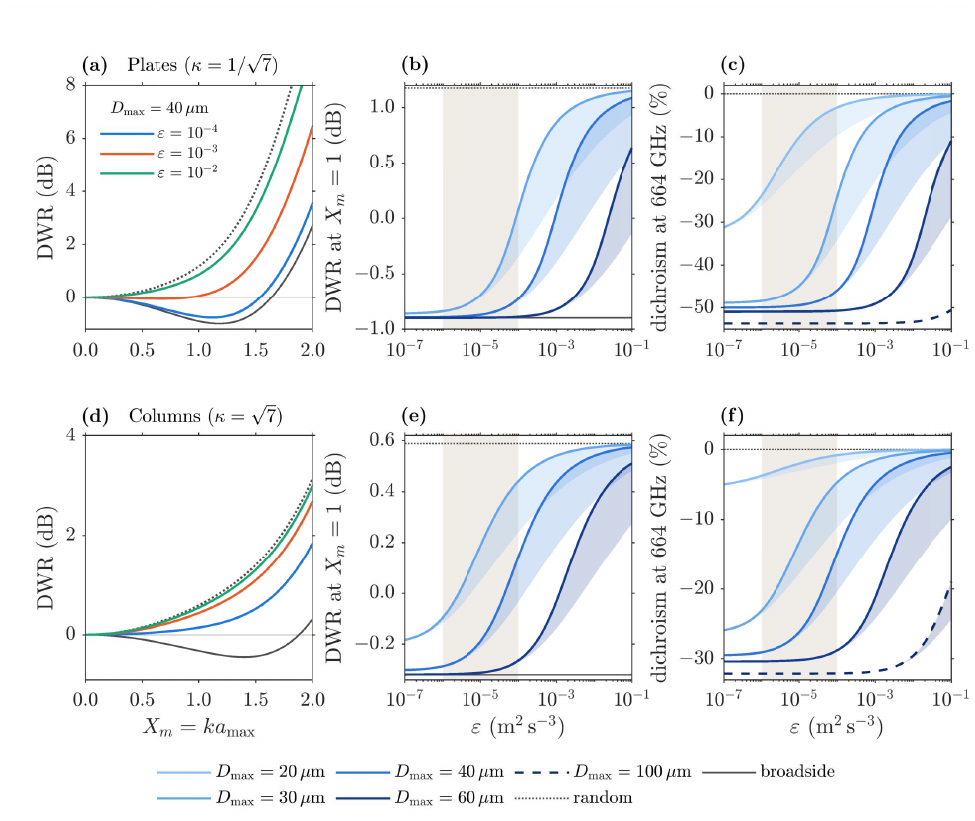}
\caption{Finite size parameter, T-matrix with $m=1.79+0.01i$, plates (top) and columns (bottom). (a,d) Dual-wavelength ratio $-10\log_{10}\rho_b$ at vertical incidence against $\Xm$ for random and broadside orientation and for $\Dmax=40~\mu$m crystals at three dissipation rates. (b,e) The same at $\Xm=1$ against $\eps$. (c,f) Extinction dichroism $(\sigma_V-\sigma_H)/\bar\sigma$ at a zenith angle of $53^\circ$ and 664~GHz, with the refractive index of \citet{warren2008optical} at that frequency. Lines and shading as in figure~\ref{fig:alignment}.}
\label{fig:finite}
\end{figure*}

At vertical incidence broadside crystals present their smallest dimension along the beam. Their backscatter then exceeds the Rayleigh value, $\rho_b>1$, up to $\Xm\approx1.6$ for plates and 1.9 for columns, so their dual-wavelength ratio is negative over this range (figure~\ref{fig:finite}). Randomly oriented crystals have $\rho_b<1$ at every $\Xm$. At $\Xm=1$ the ratio moves from $-0.90$~dB (broadside) to $+1.18$~dB (random) for plates and from $-0.32$ to $+0.59$~dB for columns. For a 40-$\mu$m plate with uniform dissipation the central 80\% of this swing spans 2.2 decades of $\eps$ (figure~\ref{fig:finite}b). A retrieval that assumes random orientation therefore interprets a partly aligned population as smaller particles and cannot fit the negative ratios of aligned crystals \citep{hogan2000measuring,matrosov2005dual,battaglia2014g}. For crystals inside the window this requires wavelengths below about 0.5~mm: at 240~GHz a 65-$\mu$m crystal has $\Xm=0.16$ and a ratio below 0.05~dB. At millimeter wavelengths the effect concerns crystals of 0.5--2~mm, outside the window, for which figure~\ref{fig:finite}a,d gives limits. The extinction dichroism at $53^\circ$ is the single-scattering source of the polarization difference measured by conical-scanning radiometers. At 664~GHz, where crystals inside the window have $\Xm\le0.45$, it is $-50\%$ for broadside plates and $-30\%$ for broadside columns, and it scales with $S$ (figure~\ref{fig:finite}c,f).

\begin{figure*}[tp]
\centering
\includegraphics[width=1\linewidth]{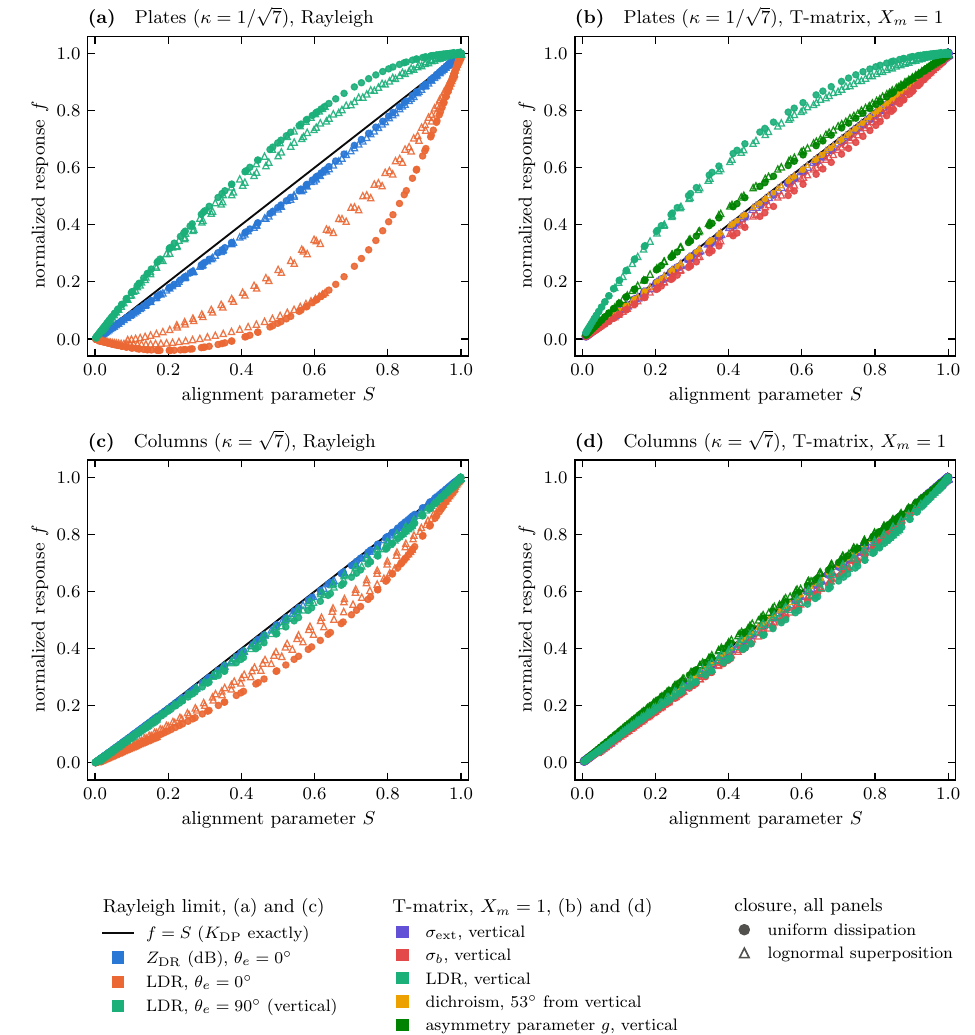}
\caption{Normalized response from equation~(\ref{eq:fQ}) of each observable against the alignment parameter for all crystal sizes and dissipation rates, plates (top) and columns (bottom). (a,c) Rayleigh limit; (b,d) T-matrix at $\Xm=1$ with $m=1.79+0.01i$. Each observable has one color in all panels; the angles are elevations $\theta_e$ of the radar beam (a,c) and the dichroism is for a line of sight $53^\circ$ from the vertical (b,d). Filled circles: uniform dissipation; open triangles: lognormal superposition. The diagonal $f=S$ is followed exactly by $\Kdp$; observables controlled by $\PL{4}$ or higher order parameters depart from it and separate the two closures.}
\label{fig:hierarchy}
\end{figure*}

Figure~\ref{fig:hierarchy} tests which observables are functions of $S$ alone. For each observable $Q$ we form the normalized response
\begin{equation}
f_Q=\frac{Q-Q^{\rm RO}}{Q^{\rm BO}-Q^{\rm RO}},
\label{eq:fQ}
\end{equation}
where RO denotes random orientation, and plot it against $S$ for every crystal size, dissipation rate and both closures. The specific differential phase lies exactly on the line $f_Q=S$. In the Rayleigh limit $\Zdr$ in decibels lies within 0.05 of it. The LDR departs by up to 0.47 and splits into separate branches for the two closures, because $\PL{4}$ and the tails of the tilt distribution control it. At $\Xm=1$ the extinction and dichroism stay within 0.02 and 0.013 of $f_Q=S$, and the backscatter and asymmetry parameter within 0.07 for plates and 0.05 and 0.03 for columns. The asymmetry parameter of solid ice departs already as $\Xm\to0$ (0.047 for plates at $\Xm=0.25$), because it is a ratio of two orientation-dependent quantities. Observables linear in the scattering amplitude, or reducing to a single projection of the polarizability, measure $S$. Cross-polar and higher-moment observables also depend on the shape of the tilt distribution.

\section{Retrieval biases}
\label{sec:bias}

\begin{figure*}[tp]
\centering
\includegraphics[width=1\linewidth]{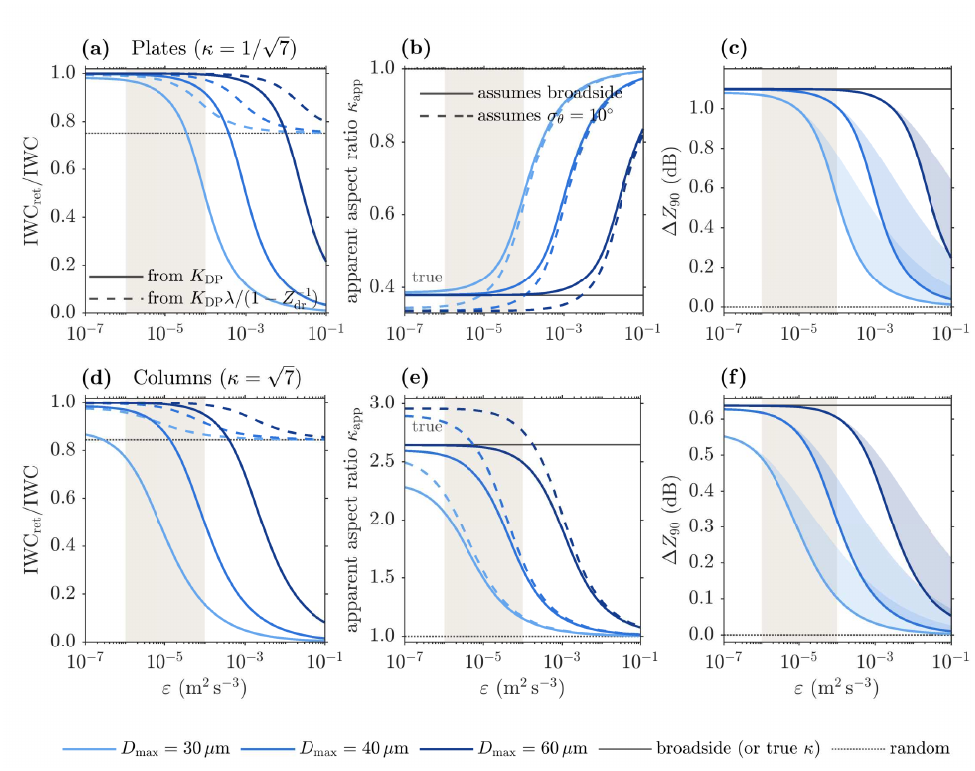}
\caption{Retrieval biases for a single crystal size, plates (top) and columns (bottom). (a,d) Ice water content retrieved from $\Kdp$ assuming broadside alignment (solid), equal to $S$, and from $\Kdp\lambda/(1-Z_{\rm dr}^{-1})$ calibrated at broadside alignment (dashed); dotted line: random-orientation limit of the latter. Size distributions are treated in table~\ref{tab:psd}. (b,e) Aspect ratio retrieved from the horizontal-incidence $\Zdr$ assuming broadside alignment (solid) or Gaussian flutter of $10^\circ$ (dashed); the solid gray line is the true value. (c,f) Excess of the vertical-incidence reflectivity over its random-orientation value. Shading in (c,f) as in figure~\ref{fig:alignment}.}
\label{fig:bias}
\end{figure*}

Equation~(\ref{eq:kdp}) inverts for the ice water content under an assumed orientation state. For a single crystal size a retrieval that assumes broadside alignment returns
\begin{equation}
\frac{\mathrm{IWC}_{\rm ret}}{\mathrm{IWC}}=\frac{S(\eps)}{S_{\rm assumed}},
\label{eq:iwcbias}
\end{equation}
exactly within the Rayleigh approximation, with $S_{\rm assumed}=1$; a Gaussian flutter of $10^\circ$ has $S_{\rm assumed}=0.913$ for plates and 0.914 for columns. The bias is therefore the alignment parameter of figure~\ref{fig:alignment}. For 40-$\mu$m plates the $\Kdp$-based ice water content is low by 8\% at $\eps=10^{-4}$ and by 50\% at $10^{-3}$~m$^2$~s$^{-3}$, and for 30-$\mu$m plates by 51\% at $10^{-4}$~m$^2$~s$^{-3}$ (figure~\ref{fig:bias}a). The $10^\circ$ assumption trades this for an overestimate of up to 9.5\% when the crystals are better aligned. $\Kdp$ itself is small for these crystals: 0.01~g~m$^{-3}$ of broadside plates gives 0.35~deg~km$^{-1}$ at 94~GHz and horizontal incidence. The same ice water content of 20--60-$\mu$m plates gives a reflectivity of only $-47$ to $-32$~dBZ. Such echoes are near the sensitivity limit of cloud radars and below that of scanning weather radars, at whose wavelengths $\Kdp$ is also 10--34 times smaller.

Because $\Kdp$ and $1-Z_{\rm dr}^{-1}$ both scale with the orientation order, their ratio has been proposed as an ice-water-content estimator insensitive to aspect ratio and canting \citep{ryzhkov2018ice,bukovcic2020polarimetric,nguyen2019determination}. For a single size, equations~(\ref{eq:addition})--(\ref{eq:kdp}) give it at horizontal incidence in closed form,
\begin{equation}
\dfrac{\Kdp\lambda}{1-Z_{\rm dr}^{-1}}=\dfrac{180\,\mathrm{IWC}}{\rho_p}\;
\dfrac{Z_{hh}}{2\ahat_\perp+\dfrac67\dah+\dfrac17\dah\,\PL{4}/\PL{2}},
\label{eq:Q}
\end{equation}
for real polarizabilities, with $Z_{hh}$ the right-hand side of equation~(\ref{eq:Z}). Orientation cancels exactly only as $\dah\to0$, the limit of low-density aggregates for which the estimator was designed. For solid ice it survives through $Z_{hh}$ and through $\PL{4}/\PL{2}$, which runs from one (broadside plates) to zero (random). Normalized to broadside alignment, the estimator falls to $\Gamma=0.751$ for random plates and 0.844 for random columns, its minimum over the states of both closures. For a single size, within these orientation families and shapes, the orientation error is therefore at most 25\% and 16\% (figure~\ref{fig:bias}a,d).

For a size distribution the cancellation is incomplete, because $\Kdp$ weights the crystals by volume $V$ and the reflectivities by $V^2$, and alignment depends strongly on size. Both are linear in the order parameters (appendix~B), so the estimator calibrated at broadside alignment returns exactly
\begin{equation}
\frac{\mathrm{IWC}_{\rm ret}}{\mathrm{IWC}}=\frac{S_K}{S_Z}\,\Gamma,
\label{eq:psd}
\end{equation}
where $S_K$ and $S_Z$ are the alignment parameters averaged with weights $n(D)V$ and $n(D)V^2$ over the number distribution $n(D)$, and $\Gamma$ is the single-size ratio of equation~(\ref{eq:Q}) evaluated with the reflectivity-weighted order parameters. $\Kdp$ alone returns $S_K$. For all the distributions computed here $0.75\le\Gamma\le1$ (plates) and $0.84\le\Gamma\le1$ (columns), so the single-size bound applies to $\Gamma$. The factor $S_K/S_Z$, however, is below one whenever alignment increases with size, and $\Kdp$ and $\Zdr$ do not constrain it. Equal ice masses of 20- and 60-$\mu$m plates at $\eps=10^{-4}$~m$^2$~s$^{-3}$, with $S=0.06$ and 0.995, have $S_K=0.53$ and $S_Z=0.96$; the estimator returns 0.55 of the ice water content, and $\Kdp$ alone 0.53. Table~\ref{tab:psd} gives both retrievals for gamma distributions $n\propto D^{\mu_D}e^{-\Lambda_DD}$ on $5\le D\le65~\mu$m, with $D_m$ the mass-weighted mean size of the truncated distribution ($\Lambda_D<0$ for $D_m=55~\mu$m and $\mu_D\le1$). Where $D_m$ is close to $D_{1/2}$, the combined estimator raises the $\Kdp$ value by factors of up to 2.5 and returns 38\%--88\% of the ice water content. Over $\mu_D=0$--6, $D_m=15$--55~$\mu$m and both closures it is within 25\% for every distribution with $D_m>1.52D_{1/2}$ (plates) or $1.44D_{1/2}$ (columns). Most of the mass is then aligned, and $\Kdp$ alone is already within 28\% (Supplement S3). This condition is sufficient, not necessary: between $D_{1/2}$ and $1.5D_{1/2}$, 41\%--51\% of the cases are within 25\%, among them plates with $D_m=40~\mu$m at $\eps=10^{-4}$~m$^2$~s$^{-3}$ ($D_m/D_{1/2}\approx1.3$; 0.77 in table~\ref{tab:psd}). $\Zdr$ does not separate the accurate from the inaccurate cases. Plate distributions with $\Zdr\ge3.5$~dB, near the broadside 4.13~dB, include cases with $\mathrm{IWC}_{\rm ret}/\mathrm{IWC}=0.51$. In these, Brownian rotation randomizes the small crystals that carry much of the mass, while the large ones, which dominate the reflectivity, are aligned. Intermittency raises the combined values by up to 0.19 but leaves this pattern (Supplement). Truncating the distributions at the edge of the window can bias the results either way (Supplement S3). Larger crystals dominate the reflectivities far more: if 2\% of the mass is in 300-$\mu$m crystals and the rest in 30-$\mu$m crystals, the large ones give about 95\% of $Z_{hh}$. $S_Z$ and $\Zdr$ then describe the large crystals, while $\Kdp$ mainly describes the small ones.

\begin{table*}[tp]
\caption{Ice water content retrieved from $\Kdp$ alone and from $\Kdp\lambda/(1-Z_{\rm dr}^{-1})$, both calibrated at broadside alignment, as fractions of the true value (entries: $\Kdp$ / combined). Gamma size distributions with $\mu_D=2$ on $5\le D\le65~\mu$m and mass-weighted mean size $D_m$; uniform dissipation with Brownian rotation; horizontal incidence. The first row gives the single-size half-alignment size.}
\label{tab:psd}
\begin{center}
\begin{tabular}{lccccccc}
\hline\hline
 & \multicolumn{3}{c}{plates, $\eps$ (m$^2$~s$^{-3}$)} & & \multicolumn{3}{c}{columns, $\eps$ (m$^2$~s$^{-3}$)}\\
\cline{2-4}\cline{6-8}
 & $10^{-6}$ & $10^{-4}$ & $10^{-2}$ & & $10^{-6}$ & $10^{-4}$ & $10^{-2}$\\
\hline
$D_{1/2}$ ($\mu$m) & 20 & 30 & 53 & & 26 & 40 & 71\\
$D_m=20~\mu$m & 0.44 / 0.52 & 0.16 / 0.30 & 0.02 / 0.18 & & 0.25 / 0.38 & 0.07 / 0.26 & 0.01 / 0.22\\
$D_m=30~\mu$m & 0.77 / 0.79 & 0.46 / 0.55 & 0.09 / 0.28 & & 0.58 / 0.65 & 0.25 / 0.42 & 0.03 / 0.31\\
$D_m=40~\mu$m & 0.92 / 0.93 & 0.73 / 0.77 & 0.23 / 0.45 & & 0.82 / 0.84 & 0.49 / 0.63 & 0.08 / 0.46\\
$D_m=50~\mu$m & 0.99 / 0.99 & 0.91 / 0.93 & 0.44 / 0.67 & & 0.95 / 0.96 & 0.74 / 0.83 & 0.16 / 0.65\\
\hline\hline
\end{tabular}
\end{center}
\end{table*}

Near random orientation the estimator is the ratio of two small measured quantities, so its model bias and its measurement error must be treated separately. An error $\sigma$ (dB) in $\Zdr$ produces a relative error $(\ln10/10)\,\sigma/(Z_{\rm dr}-1)$ in $1-Z_{\rm dr}^{-1}$. For $\sigma=0.1$~dB it is 9\% at $\Zdr=1$~dB, 19\% at 0.5~dB and 25\% at 0.38~dB, where single plates have $S=0.11$ and $\Gamma=0.77$. The single-size bias bound is therefore approached only where the noise term is as large, and the small $\Kdp$ of these crystals adds an independent error.

Shape retrievals invert $\Zdr$ for the aspect ratio under an assumed orientation state \citep{melnikov2013axis,myagkov2016relationship,matrosov2017atmospheric}. Under broadside alignment the apparent aspect ratio $\kappa_{\rm app}$ moves from the true value toward unity as $\eps$ increases (figure~\ref{fig:bias}b,e). The apparent aspect ratio thus measures turbulence as well as shape. At vertical incidence broadside plates present their largest polarizability to both polarizations, and their reflectivity exceeds the random-orientation value by 1.10~dB (columns 0.64~dB; figure~\ref{fig:bias}c,f). With the 94-GHz relation of \citet{hogan2006retrieval}, $d\log_{10}\mathrm{IWC}/dZ=0.0923+0.00058\,T$ per dBZ with $T$ in $^\circ$C, these excesses become overestimates of 17\% and 10\% at $-50^\circ$C, which turbulence removes along the curves of $S$.

\section{Specular reflection at lidar wavelengths}
\label{sec:specular}

The most direct evidence for oriented plates comes from optical wavelengths. Zenith- and nadir-pointing lidars receive strong, weakly depolarized returns from plates with their faces normal to the beam \citep{platt1978lidar,thomas1990lidar,sassen2001midlatitude}. Satellite imagers detect the corresponding glint \citep{chepfer1999observations,marshak2017terrestrial} in roughly half of the suitable cloud fields \citep{breon2004horizontally,varnai2026statistics}. Specular reflection preserves polarization, as droplets do, so these returns contaminate lidar phase classification. Tilting the CALIPSO lidar from $0.3^\circ$ to $3^\circ$ off nadir reduced the fraction of ice clouds with a significant specular contribution from about a third to about 20\% \citep{hu2009calipso,avery2020caliop,kikuchi2021climatological}. Near-nadir specular returns occur mainly between 0 and $-40^\circ$C, most often near $-15^\circ$C \citep{avery2020caliop}, and oriented crystals over Beijing most often between $-20$ and $-8^\circ$C, often beneath supercooled liquid layers \citep{westbrook2010doppler,wu2026observation}.

\subsection{Face reflection and pointing angle}

For plates with size parameters $\pi\Dmax/\lambda\gtrsim60$ at lidar wavelengths, reflection from the basal faces follows geometric optics. The hydrodynamics of section~2 treat a plate as a spheroid, whereas the reflection requires the flat basal faces of a hexagonal prism. We assume that a prism of the same maximum dimension and aspect ratio has the orientation statistics of the spheroid. This holds to the extent that their torque coefficients agree, since $D_{1/2}\propto F_b^{-1/4}$. A face of area $A_f$ whose normal is along the beam reflects a fraction $R$ of the incident power back toward the receiver. A population with a density $n_f$ of face normals per unit solid angle then contributes $A_fR_{\rm eff}\,n_f(\hat{\boldsymbol{e}}_b)/4$ to the differential backscattering cross section. Here $\hat{\boldsymbol{e}}_b$ is the beam direction and $R_{\rm eff}\simeq0.036$ is the normal-incidence reflectance of the two faces of an ice plate for $n=1.312$ at 532~nm. With $n_f=\psi(\mu)/2\pi$ on the upper hemisphere, the face term relative to random orientation is enhanced by
\begin{equation}
G(\theta_p)=\psi(\cos\theta_p)\simeq\frac{1}{\sigma_{\rm eq}^2}\exp\left(-\frac{\theta_p^2}{2\sigma_{\rm eq}^2}\right),
\label{eq:G}
\end{equation}
where $\theta_p$ is the angle of the beam from the vertical. At nadir $G=\sigma_{\rm eq}^{-2}$, which by equation~(\ref{eq:sigeq}) grows as $(W_b/u_\eta)^4$ and falls as $\eps^{-1}$ once the plates are aligned. At a fixed off-nadir angle $G$ is largest when $\sigma_{\rm eq}=\theta_p/\sqrt2$, with maximum $2/(e\theta_p^2)$, equal to 268 at $\theta_p=3^\circ$ for $\sigma_{\rm eq}=2.1^\circ$: better aligned plates send their reflection past the beam. For a thin plate, whose extinction cross section equals its face area, the face term of random plates is a fraction $f_s=R_{\rm eff}S_{\rm lid}/8\pi$ of their backscatter, with $S_{\rm lid}$ the lidar ratio. For $S_{\rm lid}=25$~sr, $f_s\approx0.036$, so the face term of a population exceeds the random backscatter for $G\gtrsim28$. The edges of the $\kappa=1/\sqrt7$ plates raise their extinction by a factor of about 1.9, which lowers $f_s$ to about 0.02 and raises this threshold to about 50. Diffraction spreads the reflection over angles of order $\lambda/\Dmax$. We convolve $\psi$ with the backscatter pattern of a circular face, $[2J_1(x)/x]^2$ with $x=k\Dmax\sin\Delta$ and $\Delta$ the angle between normal and beam, and replace the rings beyond the third zero by their mean. This lowers $G(0.3^\circ)$ by at most 15\% for $\Dmax\le60~\mu$m at 532~nm and sets a floor $G(3^\circ)\approx6$--13 for perfectly aligned plates of 60--30~$\mu$m.

These results agree with the physical-optics calculations of \citet{okamoto2020wavelength} for hexagonal plates with Gaussian tilts. For a beam $3^\circ$ off nadir both give the largest backscatter for a width of $2^\circ$ and the smallest for $0.5^\circ$. Their ratio of the backscatter for $2^\circ$ and $1^\circ$ rises from about 3.5 at an effective radius $r_{\rm eff}=10~\mu$m to 6.6 at 100~$\mu$m, and ours, through the diffraction lobe, from 3.6 at $\Dmax=20~\mu$m to 6.7 at 200~$\mu$m. We chose the mapping $\Dmax=2r_{\rm eff}$ to match this, so this part of the agreement is a calibration. The test is that $f_sG$ for $f_s=0.02$--0.036 then brackets their ratio of oriented to random backscatter for all five widths at $r_{\rm eff}=10$--30~$\mu$m. For widths of $1^\circ$--$5^\circ$ the bracket holds for any mapping up to $\Dmax=3r_{\rm eff}$.

\subsection{Dependence on dissipation rate}

\begin{figure*}[tp]
\centering
\includegraphics[width=1\linewidth]{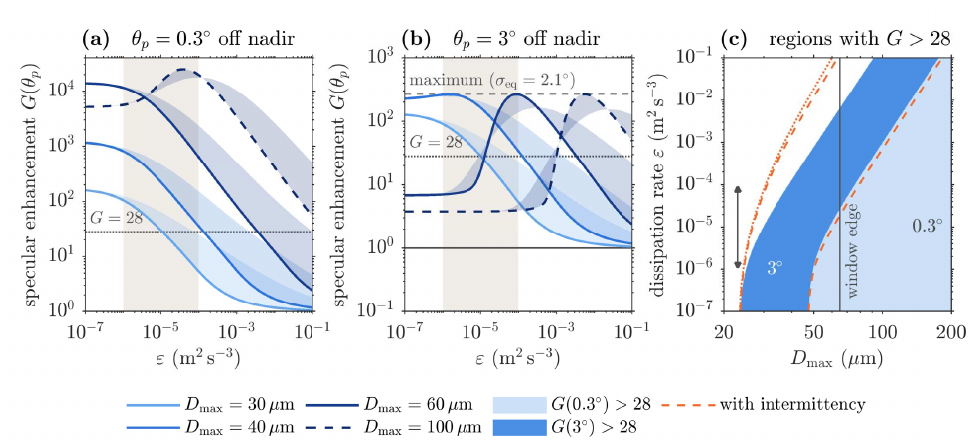}
\caption{Specular enhancement from equation~(\ref{eq:G}) of the face reflection of plates at 532~nm, including the diffraction lobe of the face, for a beam (a) $0.3^\circ$ and (b) $3^\circ$ off nadir, the two CALIPSO pointing angles. The dotted line $G=28$ marks where the face term equals the backscatter of randomly oriented thin plates with lidar ratio 25~sr (about 50 for the $\kappa=1/\sqrt7$ plates); the dashed line in (b) is the maximum $2/(e\theta_p^2)$, reached at $\sigma_{\rm eq}=\theta_p/\sqrt2$. Lines and shading as in figure~\ref{fig:alignment}; the 100-$\mu$m curve lies outside the overdamped window. (c) Sizes and dissipation rates for which $G>28$ at $0.3^\circ$ (light) and $3^\circ$ (dark) with uniform dissipation; orange lines: boundaries with the lognormal superposition. The vertical line marks the edge of the overdamped window and the arrow the upper-tropospheric range of $\eps$.}
\label{fig:specular}
\end{figure*}

Figure~\ref{fig:specular} shows $G$ at the two CALIPSO pointing angles. Near nadir (figure~\ref{fig:specular}a) $G$ exceeds 28 for plates larger than 25~$\mu$m at $\eps=10^{-6}$, 39~$\mu$m at $10^{-4}$ and 51~$\mu$m at $10^{-3}$~m$^2$~s$^{-3}$. In the aligned regime it falls by about 9~dB per decade of $\eps$, so a turbulent layer switches off the near-nadir specular return of small plates. Switch-offs observed near cloud base have been attributed both to enhanced dissipation \citep{wu2025assessment} and to the rounding of sublimating crystals \citep{westbrook2010doppler}, for plates of 0.4--1.8~mm that lie far outside the window. At $3^\circ$ off nadir (figure~\ref{fig:specular}b) $G$ is nonmonotonic in $\eps$. The widths for which $G(3^\circ)>28$, roughly $1^\circ$--$10^\circ$, belong to partly aligned plates of 25--50~$\mu$m at $\eps=10^{-6}$, 39--76~$\mu$m at $10^{-4}$ and 52--100~$\mu$m at $10^{-3}$~m$^2$~s$^{-3}$ (figure~\ref{fig:specular}c). The sizes above 65~$\mu$m extrapolate the closure beyond the overdamped window. Off-nadir pointing thus removes the specular returns of large, well-aligned plates but not those of plates near the alignment transition. This is consistent with the specular contribution that persists $3^\circ$ off nadir \citep{hu2009calipso,noel2010global,kikuchi2021climatological}, which the $3^\circ$ pointing of the EarthCARE lidar was chosen to limit \citep{wehr2023earthcare}. The maximum of $G$ falls as $\theta_p^{-2}$, to 97 at $5^\circ$ and 11 at $15^\circ$.

Intermittency matters because $G$ samples the tail of the tilt distribution. In the aligned regime the alignment time is shorter than $\tau_K$ and the static superposition applies. At nadir it multiplies $G$ by $\langle\phi^{-1}\rangle=e^{\slnphi}$ without Brownian rotation. $\LDR(90^\circ)$ carries the same factor through $\langle\phi^2\rangle$ (section~3b), but there the most turbulent regions dominate and here the quietest. Brownian rotation sets the tilt in the quietest regions and makes the factor smaller, 4.2 instead of 6.6 for 60-$\mu$m plates at $\eps=10^{-4}$~m$^2$~s$^{-3}$.

The high-spectral-resolution lidar of EarthCARE measures the lidar ratio $S_{\rm lid}$ and the particle linear depolarization ratio $\delta$ at 355~nm and $3^\circ$ off nadir \citep{wehr2023earthcare}. For such a lidar the face term adds copolar backscatter without extinction. Let $\Xi=f_s(G-1)$ be the face term in excess of its random-orientation value, relative to the random backscatter. Then $S_{\rm lid}=S_{{\rm lid},r}E/(1+\Xi)$ and $\delta=\delta_r/[1+\Xi(1+\delta_r)]$, with $S_{{\rm lid},r}$ and $\delta_r$ the random-orientation values and $E$ the extinction relative to random orientation (1.07--1.11 for the present plates, about 2 for thin plates). Eliminating $\Xi$ gives
\begin{equation}
S_{\rm lid}=\frac{S_{{\rm lid},r}E(1+\delta_r)}{\delta_r}\,\frac{\delta}{1+\delta},
\label{eq:faceline}
\end{equation}
which is linear in $\delta/(1+\delta)$, and is the locus of a fixed ratio of extinction to cross-polar backscatter, as in the physical-optics relations of \citet{okamoto2019interpretation} for this lidar. Any copolar excess at fixed extinction and cross-polar backscatter, retrieval errors included, moves a measurement along the same curve, so the position on it does not by itself identify specular reflection. Errors in the retrieved extinction, for example in the multiple-scattering correction, move it across the curves instead. Mixtures of random ice with supercooled water ($S_{\rm lid}\approx18$~sr, $\delta\approx0$) cannot reach $S_{\rm lid}<8$~sr. At 355~nm ($R_{\rm eff}=0.038$, $S_{{\rm lid},r}=30$~sr) and 568~hPa, a single size of the present plates reaches $\Xi=6.3$ ($\delta=0.036$, $S_{\rm lid}=4.4$~sr) at $\sigma_{\rm eq}=2.1^\circ$, close to the bound $\Xi\approx6.4$ set by $G(3^\circ)\le268$. Because $\sigma_{\rm eq}\propto\Dmax^{-4}$, gamma distributions with extinction-weighted mean sizes of 40--100~$\mu$m reach less: at most $\delta\approx0.10$ and $S_{\rm lid}\approx12$~sr for broad ($\mu_D=1$) and $\delta\approx0.06$ and $S_{\rm lid}\approx7$~sr for narrow ($\mu_D=20$) distributions. Lidar ratios below 8~sr at $\delta<0.06$ therefore require tilt widths of about $1.3^\circ$--$4^\circ$ within a narrow range of sizes near the alignment transition.

\subsection{EarthCARE lidar observations}

\begin{figure*}[tp]
\centering
\includegraphics[width=1\linewidth]{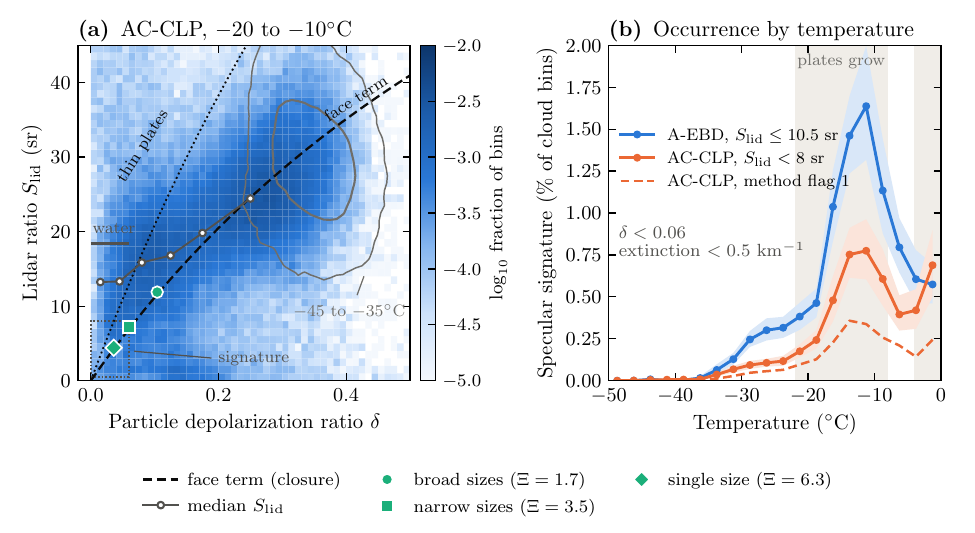}
\caption{EarthCARE lidar observations at 355~nm, $3^\circ$ off nadir, from 273 frames between December 2024 and November 2025. (a) Lidar ratio against particle depolarization ratio of low-extinction cloud bins (extinction below 0.5~km$^{-1}$) at $-20$ to $-10^\circ$C in the JAXA AC-CLP product. Open circles: median lidar ratio $S_{\rm lid}$ in bins of $\delta$. Short bar: the same for the low-extinction bins of the AC-CLP supercooled-water class with $\delta<0.06$. Gray contours enclose 50\% and 90\% of the low-extinction bins at $-45$ to $-35^\circ$C. Dashed line: equation~(\ref{eq:faceline}) for the present plates ($E=1.07$); dotted line: thin plates ($E=2$), both with the reference values of the cold bins. Markers give the largest face term $\Xi$ of the closure for the present plates with uniform dissipation and an extinction-weighted mean size of 60~$\mu$m: a broad gamma size distribution ($\mu_D=1$), a narrow one ($\mu_D=20$) and a single size. The dotted box is the specular signature. (b) Fraction of cloud bins with the specular signature in the ESA A-EBD and JAXA AC-CLP products, with 5\%--95\% ranges from a bootstrap over frames. Dashed: AC-CLP bins with method flag 1 only. Shading marks the temperatures at which plates grow \citep{bailey2009comprehensive}.}
\label{fig:earthcare}
\end{figure*}

We examined 273 frames from 22 days between December 2024 and November 2025 in the ESA A-EBD and JAXA AC-CLP products (Supplement). Low-extinction ice (extinction below 0.5~km$^{-1}$) at $-45$ to $-35^\circ$C gives $S_{{\rm lid},r}=24$~sr and $\delta_r=0.33$ in A-EBD and 30~sr and 0.35 in AC-CLP. At $-20$ to $-10^\circ$C the median $S_{\rm lid}$ of low-extinction cloud bins follows equation~(\ref{eq:faceline}) for $0.15\lesssim\delta\lesssim0.3$ but stays at 13--16~sr for $\delta<0.10$ (figure~\ref{fig:earthcare}a). In AC-CLP, 24\% of the bins with $\delta<0.06$ have $S_{\rm lid}<8$~sr, which we take as the specular signature. A-EBD lidar ratios pile up at a lower limit near 10~sr, so there we take $S_{\rm lid}\le10.5$~sr. The signature is almost absent below $-35^\circ$C and is most frequent near $-11^\circ$C, at 1.6\% (A-EBD) and 0.8\% (AC-CLP) of the cloud bins (figure~\ref{fig:earthcare}b). Plates grow at these temperatures \citep{bailey2009comprehensive}, and CALIPSO saw near-nadir specular returns most often there \citep{avery2020caliop}. AC-CLP bins with method flag 1 alone give about half this occurrence, with the maximum near $-14^\circ$C.

The analysis is exploratory. The level of the signature changes by up to an order of magnitude with the thresholds. In every sensitivity test of Supplement S7 that keeps a lidar-ratio criterion, except the AC-CLP bins with method flag 2, the maximum stays in the $2.5^\circ$C bins centered at $-11$ or $-14^\circ$C. The occurrence at $-20$ to $-10^\circ$C then exceeds that at $-35$ to $-20^\circ$C by factors of 3.3--7.8. In AC-CLP, however, about 20\%--25\% of the weakly depolarizing low-extinction bins have $S_{\rm lid}<8$~sr at all temperatures from $-45$ to $-8^\circ$C, so there the temperature dependence reflects mainly how often such layers occur. In A-EBD the corresponding fraction rises from 5\% or less below $-35^\circ$C to 17\%--19\% at $-15$ to $-10^\circ$C. The two products rarely agree on individual bins. The classifications that label these bins, mostly aerosol in the ESA lidar-only product and mostly plates in AC-CLP, derive from the same lidar signals and are not independent evidence. Finally, neither the sizes of the reflecting plates nor the dissipation rate at the scale of the signature is available. We therefore cannot test whether these are the plates near the alignment transition that the closure requires.

\section{Discussion and conclusions}
\label{sec:discussion}

Table~\ref{tab:status} classifies the results by reliability. Exact identities hold for any orientation density axisymmetric about the vertical, and therefore for any turbulence closure. Examples are the reduction of Rayleigh polarimetry to $\PL{2}$ and $\PL{4}$ and the linearity of $\Kdp$ and the polarized extinction difference in $S$. Asymptotic results require well-aligned crystals, as the LDR forms~(\ref{eq:ldr}) and the Gaussian form of $G$ do, or small size parameters, as the $O(X^2)$ forms~(\ref{eq:rgaX2}) do. The dependence on $\eps$ and size comes from the closure of section~2 and carries its uncertainties. The numerical checks (appendix~D) verify the computations, not the closure.

\begin{table*}[tp]
\caption{Status of the main results and the uncertainty that controls each. ``Exact'' means exact for any orientation density axisymmetric about the vertical within the stated scattering approximation.}
\label{tab:status}
\begin{center}
\begin{tabular}{>{\raggedright\arraybackslash}p{15.5pc}>{\raggedright\arraybackslash}p{7.5pc}>{\raggedright\arraybackslash}p{12.5pc}}
\hline\hline
Result & Status & Controlling uncertainty\\
\hline
Rayleigh observables depend on $\PL{2}$ and $\PL{4}$ only; $\Kdp/\Kdp^{\rm BO}=S$; $\sigma_V-\sigma_H\propto S$ [(\ref{eq:addition})--(\ref{eq:kdp})] & exact (Rayleigh) & none beyond the Rayleigh limit\\
$S$, $\sigma_{\rm eq}$ and $D_{1/2}$ against $\eps$ and size [(\ref{eq:b}), (\ref{eq:sigeq})--(\ref{eq:Dhalf}), tables~\ref{tab:crystals}--\ref{tab:half}] & closure & closure error (up to 15\%--16\% in $\langle\mu^2\rangle$, about 4\% in $D_{1/2}$); W against MA in the crossover; $F_b$ and habit; mass asymmetry; overdamped window\\
Combined estimator for one size, $\Gamma\ge0.75$ (plates), 0.84 (columns) [(\ref{eq:Q})] & exact form; bounds for the closure states of the two shapes & orientation family and shape; $\Zdr$ noise near random orientation\\
Combined estimator for a size distribution, $(S_K/S_Z)\Gamma$ [(\ref{eq:psd}), table~\ref{tab:psd}] & exact form for one shape; values from the closure & size distribution and the orientation--size relation\\
$\LDR(0^\circ)\propto\sigma_{\rm eq}^2$, $\LDR(90^\circ)\propto\Phi\sigma_{\rm eq}^4$ [(\ref{eq:ldr})] & asymptotic, $\sigma_{\rm eq}\lesssim3^\circ$ & tilt variance; fourth moment and intermittency\\
Rayleigh--Gans separation of turbulence, size and geometry [(\ref{eq:rholeg})] & exact (Rayleigh--Gans) & validity of Rayleigh--Gans for solid ice\\
$O(X^2)$ forms and the $54.7^\circ$ property [(\ref{eq:rgaX2})] & asymptotic, $O(X^2)$ & higher orders; internal field of solid ice (crossing near $53^\circ$)\\
T-matrix observables at $\Xm\le2$ [(\ref{eq:weights})] & numerically exact for spheroids & shape (spheroid against prism); refractive index\\
Specular enhancement and $S_{\rm lid}$--$\delta$ relation [(\ref{eq:G}), (\ref{eq:faceline})] & geometric optics with the closure & tilt width in a narrow size range; $R_{\rm eff}$, $E$ and diffraction; prism--spheroid correspondence\\
EarthCARE signature (figure~\ref{fig:earthcare}) & observation & retrieval limits, thresholds, classifications derived from the same signals\\
\hline\hline
\end{tabular}
\end{center}
\end{table*}

\subsection{Scope}

Besides the overdamped window (section~2c), which excludes the larger crystals that dominate radar reflectivity, the results rest on five approximations. First, the turbulent forcing is evaluated at a fixed orientation. Along trajectories the forcing felt is 7\%--17\% weaker in stochastic simulations \citep{roy2026orientation}, one reason the closure underestimates alignment in the crossover (section~2a). Second, crystals are spheroids of one aspect ratio. The Rayleigh operators depend on shape only through $\ahat_\parallel$ and $\ahat_\perp$, and hexagonal crystals are well represented by spheroids of matched polarizability \citep{westbrook2014rayleigh}. The hydrodynamic coefficient $F_b$ varies between 0.12 and 0.32 from moderately elongated columns to thin plates \citep{roy2026orientation}, which shifts $D_{1/2}$ only as $F_b^{-1/4}$. Natural plates, however, become thinner with size \citep{auer1970dimension}. With that law a 60-$\mu$m plate has $\kappa=0.21$ and falls at 3.0 instead of 5.3~cm~s$^{-1}$, and at $\eps=10^{-4}$~m$^2$~s$^{-3}$ its canting width is $4.9^\circ$ instead of $2.2^\circ$. $D_{1/2}$ changes by less than 10\% for $\eps\le10^{-4}$~m$^2$~s$^{-3}$ but reaches 68 instead of 53~$\mu$m at $10^{-2}$~m$^2$~s$^{-3}$, and equation~(\ref{eq:psd}) holds only approximately (Supplement S4). Third, the turbulence is homogeneous and isotropic. Mean shear, the anisotropic turbulence of stratified layers and electric fields \citep{weber2024ice} add torques that the closure does not contain \citep[see][for mean shear]{mishra2026orientation}. Fourth, the numbers are for air at 10~km. At 568~hPa and 258~K, where oriented plates are observed most often, the half-alignment sizes and the sizes with $\sigma_{\rm eq}=1^\circ$ are smaller by at most 2\% for $\eps\ge10^{-6}$~m$^2$~s$^{-3}$. Fifth, the crystals are symmetric. A center-of-mass offset in the plane of a plate adds a gravitational torque, and the plate settles at the tilt where the inertial torque balances it. At 10~km an offset of 5~nm, about $10^{-4}$ of the diameter, tilts a 40-$\mu$m plate by $10^\circ$, and the relative offset for a given tilt grows as $\Dmax^3$ (Supplement S4). Small plates therefore align only if they are nearly symmetric. In the laboratory, hexagonal ice plates of 50--250~$\mu$m falling at $Re_p=0.1$--1 were reported to fall in random orientation \citep{burgesser2018laboratory}, unlike thin disks sinking in oil \citep{katz1998subsuns}. If this holds for atmospheric crystals, an effect absent from the closure, such as asymmetry, broadens the tilts at $Re_p<1$, and the computed alignment is an upper estimate. 

\subsection{Implications for retrievals}

Four implications follow. First, the flutter width need not be prescribed. Equation~(\ref{eq:sigeq}) gives it from the broadside fall speed and the Kolmogorov velocity, with $\eps$ from the variance of the mean Doppler velocity \citep{borque2016unified}. Larger crystals dominate the reflectivity-weighted Doppler velocity, so it does not give $W_b$ of the small ones. Second, because $D_{1/2}\propto\eps^{1/8}$, the crystals whose orientation depends on turbulence occupy a narrow size range, roughly 20--50~$\mu$m. Observables weighted toward small crystals, such as submillimeter extinction in cold cirrus, retain a dependence on turbulence, and distributions that span this range can defeat the combined ice-water-content estimator. Third, $\Kdp$ or dichroism constrains $S$, the horizontal-incidence LDR constrains $\sigma_{\rm eq}^2$ where it exceeds the cross-polar isolation, and their combination the shape of the tilt distribution. Fourth, lidar phase algorithms, scattering databases and radiative-transfer models that assign one tilt distribution to all sizes \citep{klotzsche2006influence,saito2019oriented} can use the size- and turbulence-dependent $\sigma_{\rm eq}(\Dmax,\eps)$ instead (Supplement). At wavelengths below about 0.4~mm, an orientation assumption fixed independently of turbulence biases the asymmetry parameter in opposite directions for near-nadir and limb paths (section~4a). In hemispheric fluxes a quarter of the vertical-incidence effect survives.

\subsection{Conclusions}

The orientation assumed in retrievals of small ice crystals is not a property of the crystals alone. The turbulence along their fall paths sets it. Where Brownian rotation is negligible, a crystal is half aligned when it falls at about 1.7 Kolmogorov velocities, so the half-alignment size grows only as $\eps^{1/8}$. Aligned crystals cant by $\sigma_{\rm eq}\simeq C_\sigma(u_\eta/W_b)^2$, with $C_\sigma\approx2$~rad.

In the Rayleigh limit every polarimetric variable depends on the orientation density through $\PL{2}$ and $\PL{4}$ alone. An ice water content from $\Kdp$ that assumes broadside crystals is low by the mass-weighted alignment $S_K$. The estimator $\Kdp\lambda/(1-Z_{\rm dr}^{-1})$ limits this error to 25\% for one size, but for size distributions near the alignment transition it can return less than 40\% of the ice water content. Intermittency enters the depolarization of aligned plates but not $\Kdp$.

At finite size the leading orientation term of the Rayleigh--Gans cross sections and asymmetry parameter vanishes $54.7^\circ$ from the vertical for every orientation density, and for solid ice the crossing lies near $53^\circ$.

Lidars pointed $3^\circ$ off nadir, as on CALIPSO and EarthCARE, should receive their face term mainly from partly aligned plates near the alignment transition. EarthCARE data show a low-lidar-ratio, weakly depolarizing signature most frequent near $-11^\circ$C, but they do not give the size of the reflecting plates. Tilt widths of about $1^\circ$, as inferred from glints and lidar, require plates of about 50--120~$\mu$m in upper-tropospheric turbulence, and larger ones if the plates are thin. The next steps are to add particle inertia and crystal asymmetry to the closure, to test whether small natural plates align at $Re_p<1$, and to compare LDR, $\Kdp$ and specular returns with Doppler-derived dissipation rates \citep{wu2025assessment,wu2026observation}.

\acknowledgments
This work was supported by the Anusandhan National Research Foundation under project ANRF/ARG/2025/010666/ENS and by the ISRO--IIT Madras Space Technology Cell project SP26271464AMISRO008746. We thank ESA and JAXA for the EarthCARE data.

\datastatement
Python scripts that reproduce every number and table of this paper and its Supplement, and the data in every figure, are archived on Zenodo (https://doi.org/10.5281/zenodo.24345570). The archive also lists the EarthCARE frames used. The EarthCARE products ATL\_EBD\_2A and AC\_\_TC\_\_2B \citep{esa2025atlebd,esa2025actc} and AC\_\_CLP\_2B \citep{jaxa2025acclp} are available from the ESA Multi-Mission Algorithm and Analysis Platform after registration.

\appendix[A]
\appendixtitle{Hydrodynamic coefficients}

For a spheroid with semi-axes $(a_\perp,a_\perp,a_\parallel)$ define $\Delta(t)=(a_\parallel^2+t)^{1/2}(a_\perp^2+t)$ and the integrals
\begin{equation}
\begin{gathered}
\chi_0=\int_0^\infty\frac{dt}{\Delta},\\
J_\parallel=\int_0^\infty\frac{dt}{(a_\parallel^2+t)\Delta},\qquad J_\perp=\int_0^\infty\frac{dt}{(a_\perp^2+t)\Delta}.
\end{gathered}
\label{eq:A1}
\end{equation}
The settling factors relative to the volume-equivalent sphere and the resistance to rotation about a transverse axis are \citep{kim2013microhydrodynamics}
\begin{equation}
\begin{aligned}
\hat U_\parallel&=\frac{3R_{\rm eq}}{8}\left(\chi_0+a_\parallel^2J_\parallel\right),\\
\hat U_\perp&=\frac{3R_{\rm eq}}{8}\left(\chi_0+a_\perp^2J_\perp\right),\\
C_r&=\frac{16\pi\mu_f(a_\perp^2+a_\parallel^2)}{3(a_\perp^2J_\perp+a_\parallel^2J_\parallel)}.
\end{aligned}
\label{eq:A2}
\end{equation}
The inertial torque on a spheroid in a uniform slip $\boldsymbol{W}$ is $s_\kappa F_D\rho_f W^2a_{\max}^3(\hat{\boldsymbol{W}}\cdot\pv)(\hat{\boldsymbol{W}}\times\pv)$ at leading order in the particle Reynolds number, with $F_D(\kappa)$ given in closed form by \citet{dabade2015effects} and normalized on the largest semi-axis. Dividing by $C_r$ gives the bare coefficient $F_0=F_D\mu_fa_{\max}^3/C_r$, and with the quasi-static slip $\boldsymbol{W}=W_{\rm eq}[\hat U_\perp\gh+(\hat U_\parallel-\hat U_\perp)\mu\pv]$ the drift in (\ref{eq:pdot}) has rate (\ref{eq:lamr}) with
\begin{equation}
\begin{gathered}
F_{\rm eq}=F_0\hat U_\parallel\hat U_\perp,\\
F_b=F_0\frac{\hat U_\parallel}{\hat U_\perp}\ \text{(columns)},\qquad F_b=F_0\frac{\hat U_\perp}{\hat U_\parallel}\ \text{(plates)}.
\end{gathered}
\label{eq:A3}
\end{equation}
The orientation dependence of the slip changes only this rescaling and leaves the potential quadratic in $\mu$ \citep{roy2026orientation}. Table~\ref{tab:coef} lists the coefficients; our implementation reproduces the values tabulated by \citet{roy2026orientation} for $\kappa=2$, 3, 4, 5, 10, 0.05 and 0.03 to the quoted three digits.

\begin{table*}[tp]
\caption{Hydrodynamic coefficients of the two shapes (appendix~A).}
\label{tab:coef}
\begin{center}
\begin{tabular}{lrrrrrrr}
\hline\hline
 & $\mathcal{B}$ & $\hat U_\parallel$ & $\hat U_\perp$ & $F_D$ & $F_0$ & $F_b$ & $F_{\rm eq}$\\
\hline
plates, $\kappa=1/\sqrt7$ & $-0.75$ & 0.817 & 0.978 & 1.871 & 0.147 & 0.176 & 0.117\\
columns, $\kappa=\sqrt7$ & 0.75 & 1.037 & 0.861 & 0.886 & 0.122 & 0.147 & 0.109\\
\hline\hline
\end{tabular}
\end{center}
\end{table*}

\appendix[B]
\appendixtitle{Rayleigh orientation averages}

Let $\pv=(\sin\theta\cos\zeta,\sin\theta\sin\zeta,\cos\theta)$ with $\mu=\cos\theta$, the azimuth $\zeta$ uniform, and a beam along $(\cos\theta_e,0,\sin\theta_e)$ with $\boldsymbol{h}=(0,1,0)$ and $\boldsymbol{v}=(-\sin\theta_e,0,\cos\theta_e)$. Writing $m_2=\langle\mu^2\rangle$, $m_4=\langle\mu^4\rangle$, $s=\sin\theta_e$ and $c=\cos\theta_e$, the averages needed in (\ref{eq:Z}) are
\begin{equation}
\begin{aligned}
\langle(\pv\cdot\boldsymbol{h})^2\rangle&=\tfrac12(1-m_2),\\
\langle(\pv\cdot\boldsymbol{h})^4\rangle&=\tfrac38(1-2m_2+m_4),\\
\langle(\pv\cdot\boldsymbol{v})^2\rangle&=\tfrac12s^2(1-m_2)+c^2m_2,\\
\langle(\pv\cdot\boldsymbol{v})^4\rangle&=\tfrac38s^4(1-2m_2+m_4)\\
&\quad+3s^2c^2(m_2-m_4)+c^4m_4,\\
\langle(\pv\cdot\boldsymbol{h})^2(\pv\cdot\boldsymbol{v})^2\rangle&=\tfrac18s^2(1-2m_2+m_4)\\
&\quad+\tfrac12c^2(m_2-m_4).
\end{aligned}
\label{eq:B1}
\end{equation}
With $m_2=(1+2\PL{2})/3$ and $m_4=(7+20\PL{2}+8\PL{4})/35$ these reduce to (\ref{eq:addition}) and to
\begin{equation}
\begin{aligned}
&\langle(\pv\cdot\boldsymbol{h})^2(\pv\cdot\boldsymbol{v})^2\rangle\\
&\enspace=\frac{2s^2(7-10\PL{2}+3\PL{4})+c^2(14+10\PL{2}-24\PL{4})}{210},
\end{aligned}
\label{eq:B2}
\end{equation}
which equals $1/15$ for random orientation. At horizontal incidence $Z_{hh}-Z_{vv}\propto-\PL{2}\,[2\,\mathrm{Re}(\ahat_\perp^\ast\dah)+\tfrac67|\dah|^2]-\tfrac17|\dah|^2\PL{4}$, which with (\ref{eq:kdp}) gives (\ref{eq:Q}). For broadside plates (\ref{eq:Q}) reduces to $|\ahat_\perp|^2/(\ahat_\perp+\ahat_\parallel)$ and for random orientation, where $\PL{4}/\PL{2}\to0$, to $[\ahat_\perp^2+\tfrac23\ahat_\perp\dah+\tfrac15\dah^2]/(2\ahat_\perp+\tfrac67\dah)$.

For a distribution $n(D)$ of crystals of one shape, $\Kdp\propto\sum nV\PL{2}_D$ and $Z_{hh},Z_{vv}\propto\sum nV^2Z_{hh,vv}(\PL{2}_D,\PL{4}_D)$, where the subscript $D$ marks the order parameters of size $D$. Because $Z_{hh}$ and $Z_{vv}$ are linear in the order parameters, the distribution has the reflectivities of a single crystal with the reflectivity-weighted order parameters $\PL{L}_Z=\sum nV^2\PL{L}_D/\sum nV^2$, and the $\Kdp$ of one with the mass-weighted $\PL{2}_K$. Dividing (\ref{eq:Q}) for these two states gives (\ref{eq:psd}), with $S_K/S_Z=\PL{2}_K/\PL{2}_Z$.

In the aligned limit write the local concentration as $1/b(\phi)=A\phi+C$, with turbulent part $A=2\sigma_f^2\tau_c/[\lambda_r(1+\lambda_r\tau_c)]$ and Brownian part $C=2D_B/\lambda_r$. A Boltzmann density of large $b$ gives $\langle1-\mu^2\rangle=1/b$ and $\langle(1-\mu^2)^2\rangle=2/b^2$ for plates, so the lognormal superposition gives $\langle1-\mu^2\rangle=A+C$ and $\langle(1-\mu^2)^2\rangle=2(A^2e^{\slnphi}+2AC+C^2)$; with $\sigma_{\rm eq}^2=(A+C)/2$ these give (\ref{eq:ldr}).

\appendix[C]
\appendixtitle{Rayleigh--Gans orientation averages}

Let $s=\sin(\vartheta/2)$ and $c=\cos\vartheta$. Expanding (\ref{eq:formfactor}) gives $\mathcal{R}=1-\tfrac45X^2s^2(1+r\mu_q^2)+O(X^4)$. The scattering vector makes the angle $\beta$ with the incident direction, $\cos\beta=\hat{\boldsymbol{q}}\cdot\hat{\boldsymbol{k}}_i=-s$, so that averaging $\langle\mu_q^2\rangle=\tfrac13+\tfrac23\PL{2}P_2(\gh\cdot\hat{\boldsymbol{q}})$ over the azimuth $\xi$ of the scattered direction about the beam replaces $P_2(\gh\cdot\hat{\boldsymbol{q}})$ by $P_2(\cos\theta_i)P_2(s)$, by the addition theorem. The azimuthally averaged form factor is then
\begin{equation}
\begin{aligned}
\langle\mathcal{R}\rangle_\xi={}&1-\frac{4X^2s^2}{5}\left[1+\frac{r}{3}+\frac{2r}{3}\PL{2}P_2(\cos\theta_i)P_2(s)\right]\\
&+O(X^4).
\end{aligned}
\label{eq:C1}
\end{equation}
For unpolarized incidence the angular weight of the scattered intensity is $\mathcal{W}(c)=\tfrac38(1+c^2)$, normalized so that $\int_{-1}^1\mathcal{W}\,dc=1$, and $\sigma_s/\sigma_0=\int_{-1}^1\mathcal{W}\langle\mathcal{R}\rangle_\xi\,dc$ and $g\,\sigma_s/\sigma_0=\int_{-1}^1\mathcal{W}c\langle\mathcal{R}\rangle_\xi\,dc$. The four integrals
\begin{equation}
\begin{gathered}
\int_{-1}^{1}\mathcal{W}s^2\,dc=\frac12,\qquad\int_{-1}^{1}\mathcal{W}s^2P_2(s)\,dc=\frac{11}{40},\\
\int_{-1}^{1}\mathcal{W}cs^2\,dc=\int_{-1}^{1}\mathcal{W}cs^2P_2(s)\,dc=-\frac15
\end{gathered}
\label{eq:C2}
\end{equation}
give the first two lines of (\ref{eq:rgaX2}); in the asymmetry parameter the denominator $\sigma_s/\sigma_0$ contributes only at $O(X^4)$, because the numerator vanishes at $X=0$. At $\vartheta=\pi$, $s=1$ and $\hat{\boldsymbol{q}}=-\hat{\boldsymbol{k}}_i$, so no azimuthal average is needed and (\ref{eq:C1}) gives the third line. At $O(X^4)$ the expansion of $\mathcal{R}$ contains $r^2\mu_q^4$, which brings in $\PL{4}$; the linearity in $\PL{2}$ is a property of the leading correction only.

\appendix[D]
\appendixtitle{Numerical methods}

Moments of the Boltzmann density were evaluated in closed form with the error and Dawson functions and independently by Gauss--Legendre quadrature graded toward the attractor; the two agree to $10^{-10}$ for $10^{-6}\le b\le10^5$. The superposition (\ref{eq:mixture}) used 120-point Gauss--Hermite quadrature in $\ln\phi$, which changes the moments by less than $10^{-6}$ when doubled. The Rayleigh averages (\ref{eq:B1}) agree with direct quadrature to $10^{-13}$, and with T-matrix calculations at a volume-equivalent size parameter of 0.02 to 0.0011~dB in $\Zdr$. The T-matrix code \citep{mishchenko1998capabilities,leinonen2014high} reproduces Mie theory for a sphere to seven digits and approaches the Rayleigh--Gans average (\ref{eq:rholeg}) as $m\to1$; the truncation orders were $N=5$--7 at $\Xm\le2$. The size-distribution retrievals of table~\ref{tab:psd} reproduce the decomposition (\ref{eq:psd}) to $10^{-13}$. The Supplement gives the remaining details, including the azimuthal and angular quadratures, the specular calculations and the refractive-index tests. These checks verify the numerical evaluation of the model; the accuracy of the closure itself is that established against stochastic simulations by \citet{roy2026orientation} (section~2a). The scripts were written and run, and the text edited, with Claude Opus 5 to 5.5 (Anthropic); the authors checked the results and are responsible for the content.


\twocolumn[{%
\begin{center}
{\large\bf Supplemental Material}\par
\vskip 8pt
{\normalsize\bf Scattering by Small Ice Crystals in Turbulence: From Orientation Statistics to Retrieval Biases}\par
\vskip 8pt
{\normalsize\sc Himanshu Mishra, Mohd.\ Meraj Khan, T. Narayana Rao, and Anubhab Roy}\par
\end{center}
\vskip 14pt}]
\setcounter{section}{0}
\setcounter{subsection}{0}
\setcounter{equation}{0}
\setcounter{table}{0}
\setcounter{figure}{0}
\renewcommand\thesection{S\arabic{section}}
\renewcommand\theequation{S\arabic{equation}}
\renewcommand\thetable{S\arabic{table}}
\renewcommand\thefigure{S\arabic{figure}}

\noindent Equation, figure and table numbers without the prefix S refer to the main text. Sections~S1--S3 give the time scales behind the overdamped window, the expansion about the attractor and the size-distribution retrievals. Section~S4 compares the results with observed tilt widths and aerodynamic tilt models and treats crystal habit and asymmetry. Section~S5 discusses radiative-transfer studies, section~S6 the numerical methods, and section~S7 the EarthCARE data, selection and sensitivity tests.

\section{Time scales of the overdamped window}

\citet{gustavsson2021effect} describe the angular dynamics of a small crystal settling in turbulence through three time scales: the alignment time $1/\lambda_r$ of the fluid-inertia torque, the time $\ell\eta/W_b$ in which the crystal settles through a correlation length $\ell\eta$ of the velocity gradients ($\ell=10$ in their phase diagrams), and the damping times of translation and rotation. The overdamped equation~(2), their regime~2, requires both damping times to be short compared with $\min(1/\lambda_r,\ell\eta/W_b)$. Table~S1 gives the times for the two shapes at 10~km: $\tau_p=W_b\rho_p/[g(\rho_p-\rho_f)]\simeq W_b/g$ is the response time for broadside settling, and $\tau_{\rm rot}=I_\perp/C_r$, with $I_\perp=m(a_\perp^2+a_\parallel^2)/5$ the moment of inertia about a transverse axis and $C_r$ the rotational resistance of appendix~A. Table~S2 gives the flow time scales for the largest crystals of the window.

When $\tau_p$ is finite, the slip velocity lags the fluid velocity along the fall path, its direction tilts by about $\tau_p$ times the velocity gradient, and the crystal aligns with the tilted slip. \citet{gustavsson2021effect} give the general tilt variance (their eq.~23) and its two asymptotes, which in Kolmogorov units are
\begin{equation}
\begin{gathered}
\langle\delta\varphi^2\rangle_2=f\,\frac{\langle O_{12}^2\rangle+\mathcal{B}^2\langle S_{12}^2\rangle}{\Lambda^2},\\
\langle\delta\varphi^2\rangle_3=f\left(\frac{\tau_p}{\tau_K}\right)^2\langle A_{21}^2\rangle,
\end{gathered}
\end{equation}
with $f=1$ for columns and 2 for plates, $\Lambda=\lambda_r\tau_K$, and $O_{12}$, $S_{12}$ and $A_{21}$ components of the vorticity, strain and velocity-gradient tensors; the first is the aligned limit of equation~(7). With the isotropic values $\langle O_{12}^2\rangle=1/12$, $\langle S_{12}^2\rangle=1/20$ and $\langle A_{21}^2\rangle=2/15$ their ratio is
\begin{equation}
\begin{gathered}
\frac{\langle\delta\varphi^2\rangle_3}{\langle\delta\varphi^2\rangle_2}=x^2\,\frac{2/15}{1/12+\mathcal{B}^2/20}=1.20\,x^2,\\
x=\tau_p\lambda_r\quad(|\mathcal{B}|=0.75),
\end{gathered}
\end{equation}
which places the transition between the two regimes near $x\approx0.9$. Their general result also contains a cross-correlation between the slip fluctuation and the Jeffery forcing, which is of first order in $x$. To evaluate it we use a linear model near the attractor. With gravity along $-\hat{\boldsymbol{y}}$ and $A_{ij}=\partial u_i/\partial x_j$, the tilt $\varphi$ of a plate whose axis is near $\hat{\boldsymbol{y}}$ obeys
\begin{equation}
\begin{gathered}
\dot\varphi=-\lambda_r(\varphi-\varphi_*)+F,\qquad F=O_{xy}+\mathcal{B}S_{xy},\\
\varphi_*=-\frac{w_x}{W_b},\qquad \tau_p\dot w_x=-w_x+\tau_pW_bA_{xy},
\end{gathered}
\end{equation}
where $w_x$ is the horizontal slip, which the crystal acquires by lagging the horizontal velocity of the fluid it falls through; for a column whose axis is near $\hat{\boldsymbol{x}}$, $F=-O_{xy}+\mathcal{B}S_{xy}$ and $\varphi_*=+w_x/W_b$. Taking $A_{xy}$ and $A_{yx}$ as Ornstein--Uhlenbeck processes with correlation time $\tau_c$ and the isotropic covariances $\langle A_{xy}^2\rangle=\langle A_{yx}^2\rangle=2/15$ and $\langle A_{xy}A_{yx}\rangle=-1/30$, the stationary covariance of (S3) gives, for both shapes and $\lambda_r\tau_c=10$--100,
\begin{equation}
\frac{\langle\varphi^2\rangle}{\langle\varphi^2\rangle_{\tau_p=0}}=1-0.82\,x+1.20\,x^2,
\end{equation}
to within 2\% for $x\le1$. The quadratic term is (S2). The linear term, $-2(1/12-|\mathcal{B}|/20)/(1/12+\mathcal{B}^2/20)=-0.82$, arises because the slip tilt opposes the rotation by the same velocity gradient. In a shear $u_x=\gamma y$ a falling plate rotates its axis toward $+\hat{\boldsymbol{x}}$ at the rate $(1+\mathcal{B})\gamma/2$, whereas its lag tilts the slip, and with it the attractor, toward $-\hat{\boldsymbol{x}}$ by $\tau_p\gamma$.

At $\Dmax=65~\mu$m, $x=0.12$ for plates and 0.017 for columns (table~S1), so (S4) lowers the tilt variance by 8\% and 1.4\%. It lowers it by 13\% for 80-$\mu$m plates ($x=0.42$), returns to the overdamped value at $x\approx0.69$ ($\Dmax=87~\mu$m for plates and 121~$\mu$m for columns), and for 100-$\mu$m plates ($x=1.6$), beyond the range of (S4), the linear model gives a factor of 2.6--2.7. The linear model neglects the dependence of the torque on the slip magnitude, rotational inertia and the structure of the flow along the trajectory, so (S4) estimates the size of the correction rather than replacing finite-inertia calculations. The linearized tilt equation $I_\perp\ddot\theta+C_r\dot\theta+C_r\lambda_r\theta=0$ is overdamped for $\tau_{\rm rot}\lambda_r<1/4$, which holds up to $\Dmax=90~\mu$m (plates) and 121~$\mu$m (columns). At $\Dmax=65~\mu$m the particle Reynolds number on $W_b$ and $\Dmax$ is 0.11 (plates) and 0.06 (columns); the Stokes number $\tau_p/\tau_K$ is below 0.11 for $\eps\le10^{-2}$~m$^2$~s$^{-3}$ but reaches 0.34 (plates) and 0.19 (columns) at $10^{-1}$~m$^2$~s$^{-3}$; and $\tau_pW_b/(\ell\eta)\le0.05$ (table~S2). The parameter map of \citet{gustavsson2021effect}, for platelets in air with $\nu=0.1$~cm$^2$~s$^{-1}$ and $\rho_p/\rho_f=1000$, confines overdamped dynamics to $a\beta^{1/2}\ll18~\mu$m in their notation, $(a_\perp a_\parallel)^{1/2}$ in ours. That boundary corresponds to $x\approx1$; in the air at 10~km the same condition gives $(a_\perp a_\parallel)^{1/2}\approx28~\mu$m, or $\Dmax\approx92~\mu$m for the present plates, and the window $\Dmax\le65~\mu$m, where $x\le0.12$, lies well inside it. The correlation length $\ell\eta$ is the Taylor microscale in their comparison with direct simulations ($\ell\approx15$ at $\Rey\approx56$) and is larger at atmospheric Reynolds numbers, which only weakens the conditions that involve $\ell\eta/W_b$.

Table~S2 also shows why the forcing correlation time of the closure is $\tau_K$. In the crossover, $\Sv_b\approx1.7$, a crystal settles through $\ell\eta$ in about $6\tau_K$, so the velocity gradient it samples decorrelates on $\tau_K$. Where settling is faster, $\Sv_b>\ell$, the time $\ell\eta/W_b$ is shorter than $\tau_K$, but there $\Lambda\gtrsim15$ and the concentration~(7) reduces to its quasi-static limit $30\Lambda^2/(5+3\mathcal{B}^2)$, which does not depend on the correlation time.

\begin{table*}[tp]
\caption{Particle time scales at 10~km: broadside fall speed $W_b$, response times $\tau_p$ and $\tau_{\rm rot}$, alignment time $1/\lambda_r$, their ratios, and the particle Reynolds number $Re_p$ on $W_b$ and $\Dmax$.}
\begin{center}
\begin{tabular}{rrrrrrrr}
\hline\hline
$\Dmax$ ($\mu$m) & $W_b$ (cm~s$^{-1}$) & $\tau_p$ (ms) & $\tau_{\rm rot}$ (ms) & $1/\lambda_r$ (ms) & $\tau_p\lambda_r$ & $\tau_{\rm rot}\lambda_r$ & $Re_p$\\
\hline
\multicolumn{8}{l}{plates, $\kappa=1/\sqrt7$}\\
20 & 0.585 & 0.597 & 0.179 & 5835 & $1.0\times10^{-4}$ & $3.1\times10^{-5}$ & 0.0033\\
40 & 2.34 & 2.39 & 0.716 & 365 & 0.0065 & 0.0020 & 0.027\\
60 & 5.27 & 5.37 & 1.61 & 72.0 & 0.075 & 0.022 & 0.090\\
65 & 6.18 & 6.31 & 1.89 & 52.3 & 0.12 & 0.036 & 0.11\\
80 & 9.37 & 9.55 & 2.87 & 22.8 & 0.42 & 0.13 & 0.21\\
100 & 14.6 & 14.9 & 4.48 & 9.34 & 1.6 & 0.48 & 0.42\\
150 & 32.9 & 33.6 & 10.1 & 1.84 & 18 & 5.5 & 1.4\\
\multicolumn{8}{l}{columns, $\kappa=\sqrt7$}\\
20 & 0.323 & 0.329 & 0.119 & 22969 & $1.4\times10^{-5}$ & $5.2\times10^{-6}$ & 0.0018\\
40 & 1.29 & 1.32 & 0.475 & 1436 & $9.2\times10^{-4}$ & $3.3\times10^{-4}$ & 0.015\\
60 & 2.90 & 2.96 & 1.07 & 284 & 0.010 & 0.0038 & 0.049\\
65 & 3.41 & 3.47 & 1.26 & 206 & 0.017 & 0.0061 & 0.063\\
80 & 5.16 & 5.26 & 1.90 & 89.7 & 0.059 & 0.021 & 0.12\\
100 & 8.06 & 8.22 & 2.97 & 36.8 & 0.22 & 0.081 & 0.23\\
150 & 18.1 & 18.5 & 6.69 & 7.26 & 2.5 & 0.92 & 0.77\\
\hline\hline
\end{tabular}
\end{center}
\end{table*}

\begin{table*}[tp]
\caption{Flow time scales at 10~km and the crystals of $\Dmax=65~\mu$m: Kolmogorov time $\tau_K$ and length $\eta$, the time $\ell\eta/W_b$ to settle through ten Kolmogorov lengths, the broadside settling number $\Sv_b$, $\Lambda=\lambda_r\tau_K$, the Stokes number $\tau_p/\tau_K$, and $\tau_pW_b/(\ell\eta)$.}
\begin{center}
\begin{tabular}{llrrrrrrr}
\hline\hline
shape & $\eps$ (m$^2$~s$^{-3}$) & $\tau_K$ (ms) & $\eta$ (mm) & $\ell\eta/W_b$ (ms) & $\Sv_b$ & $\Lambda$ & $\tau_p/\tau_K$ & $\tau_pW_b/(\ell\eta)$\\
\hline
plates & $10^{-6}$ & 5935 & 14.5 & 2338 & 25.4 & 113 & 0.0011 & 0.0027\\
plates & $10^{-4}$ & 593 & 4.57 & 739 & 8.03 & 11.3 & 0.011 & 0.0085\\
plates & $10^{-2}$ & 59.3 & 1.45 & 234 & 2.54 & 1.13 & 0.11 & 0.027\\
plates & $10^{-1}$ & 18.8 & 0.813 & 131 & 1.43 & 0.359 & 0.34 & 0.048\\
columns & $10^{-6}$ & 5935 & 14.5 & 4243 & 14.0 & 28.8 & $5.9\times10^{-4}$ & $8.2\times10^{-4}$\\
columns & $10^{-4}$ & 593 & 4.57 & 1342 & 4.42 & 2.88 & 0.0059 & 0.0026\\
columns & $10^{-2}$ & 59.3 & 1.45 & 424 & 1.40 & 0.288 & 0.059 & 0.0082\\
columns & $10^{-1}$ & 18.8 & 0.813 & 239 & 0.787 & 0.0912 & 0.19 & 0.015\\
\hline\hline
\end{tabular}
\end{center}
\end{table*}

\section{Expansion about the attractor}

Near alignment let $w=1-\mu^2$ for plates and $w=\mu^2$ for columns, the squared sine of the tilt from the attractor. Every orientation-averaged quantity is the average of a kernel that, after the azimuthal average, is a smooth function of $\mu^2$ and hence of $w$, so that
\begin{equation}
Q=Q^{\rm BO}+Q_1\langle w\rangle+\tfrac12Q_2\langle w^2\rangle+O(\langle w^3\rangle),
\end{equation}
with $Q_1$ and $Q_2$ the first two derivatives of the kernel at the attractor. For a ratio of two averages, such as $g$ or the LDR, $Q_1=(N_1-Q^{\rm BO}D_1)/D^{\rm BO}$ in terms of the coefficients $N_1$ and $D_1$ of numerator and denominator, and a term in $\langle w\rangle^2$ appears at second order. For a Boltzmann density of large $b$, $\langle w\rangle=b^{-1}+\tfrac12b^{-2}$ and $\langle w^2\rangle=2b^{-2}$ for plates, with errors of relative order $b^{-1}$, and $\langle w\rangle=\tfrac12b^{-1}$ and $\langle w^2\rangle=\tfrac34b^{-2}$ for columns, with exponentially small errors. In the lognormal superposition with $1/b(\phi)=A\phi+C$ (appendix~B) these moments are averaged over $\phi$: $\langle b^{-1}\rangle=A+C=2\sigma_{\rm eq}^2$ is independent of the intermittency, whereas $\langle b^{-2}\rangle=(A+C)^2\Phi$ carries it through the factor $\Phi$ of equation~(18), and the term in $\langle w\rangle^2$ carries no $\Phi$. Every broadside quantity with $Q_1\neq0$ therefore responds to the tilt at first order through $\langle w\rangle$, which is $2\sigma_{\rm eq}^2$ for plates and $\sigma_{\rm eq}^2$ for columns, and to intermittency only at second order, through $\Phi\sigma_{\rm eq}^4$. A kernel that starts at $w^2$, as the vertical-incidence cross-polar return of plates does, responds as $\Phi\sigma_{\rm eq}^4$ at leading order. The expansion is asymptotic. After the average over $\phi$ its $n$th term carries $\langle\phi^n\rangle=\exp[n(n-1)\slnphi/2]$, so the ordering requires $b\gg e^{2\slnphi}$, which is 19--100 over the range of $\eps$ considered here, that is $\sigma_{\rm eq}\lesssim3^\circ$. For less aligned crystals the turbulent tail of $P(\phi)$ contributes at first order.

At $O(X^2)$ the Rayleigh--Gans kernels are linear in $\mu^2$, so that $Q_2=0$. With $\PL{2}=1-\tfrac32\langle w\rangle$ for plates and $-\tfrac12+\tfrac32\langle w\rangle$ for columns, equation~(21) at vertical incidence gives $Q_1=\pm\tfrac{11}{50}rX^2$ for $\sigma_s/\sigma_0$, $\mp\tfrac{4}{25}rX^2$ for $g$ and $\pm\tfrac45rX^2$ for $\sigma_b/\sigma_{b0}$, with the upper signs for plates. For columns $\langle w\rangle$ is independent of the intermittency at every order of the expansion, so that intermittency enters these quantities only at $O(X^4)$, where $Q_2\neq0$. For plates it enters at $O(X^2)$ through the term $\tfrac12\langle b^{-2}\rangle$ of $\langle w\rangle$.

\section{Size-distribution retrievals}

The calculations of section~5 and table~3 use gamma number distributions $n(D)\propto D^{\mu_D}\exp(-\Lambda_DD)$ on 121 sizes $5\le D\le65~\mu$m, with $\Lambda_D$ set by the mass-weighted mean size of the truncated distribution, $D_m=\sum nD^4/\sum nD^3$. They use 61 dissipation rates $10^{-7}\le\eps\le10^{-1}$~m$^2$~s$^{-3}$, air at 10~km and horizontal incidence at 94~GHz. For $D_m=55~\mu$m this requires $\Lambda_D=-0.029$ and $-0.008~\mu$m$^{-1}$ at $\mu_D=0$ and 1, so that $n$ increases with $D$. Excluding these two distributions leaves the threshold ratios below unchanged. For one shape, $\Kdp$ weights the crystals by $nV$ and the reflectivities by $nV^2$, and both are linear in the order parameters, so the estimator returns
\begin{equation}
\frac{\mathrm{IWC}_{\rm ret}}{\mathrm{IWC}}=\frac{S_K}{S_Z}\,\Gamma\bigl(\PL{2}_Z,\PL{4}_Z\bigr),
\end{equation}
equation~(26), with $\Gamma$ the single-size ratio of equation~(25) normalized to broadside alignment. The numerical retrievals reproduce this decomposition to $10^{-13}$. Over $\mu_D=0$--6, $D_m=15$--55~$\mu$m and all dissipation rates, $\Gamma$ stays between 0.752 and 1 for plates and between 0.844 and 1 for columns, whereas $S_K/S_Z$ ranges from 0.16 to 1 for plates and from 0.17 to 1 for columns. In the two-size example of section~5, equal ice masses of 20- and 60-$\mu$m plates at $\eps=10^{-4}$~m$^2$~s$^{-3}$ have $S=0.064$ and 0.995, $S_K=0.53$, $S_Z=0.96$, $\Zdr=3.93$~dB and $\mathrm{IWC}_{\rm ret}/\mathrm{IWC}=0.545$, against 0.53 from $\Kdp$ alone. If 80-$\mu$m plates, taken as aligned, carry 10\% of the total mass, the combined value rises to 0.588. If they take that mass from the 60-$\mu$m plates it falls to 0.542, and if they take it from the 20-$\mu$m plates it rises to 0.634. Truncating the distributions at the edge of the window therefore biases the results in a direction set by the crystals beyond it.

Table~S3 extends table~3 to $\mu_D=0$ and 6 and to the lognormal superposition. Narrower distributions raise the combined values at small $D_m$, but the pattern is the same: the combined estimator differs little from $\Kdp$ alone where both are accurate and remains far below one in the crossover. Over all the cases computed, the superposition raises the combined values by up to 0.19 (plates) and 0.15 (columns). Table~S4 sorts the cases with uniform dissipation by $D_m/D_{1/2}$, with $D_{1/2}(\eps)$ the single-size half-alignment size of the same closure. The ratio above which every case is accurate is an extreme over the cases sampled. We therefore evaluate it on a finer grid, with $D_m$ in steps of 0.25~$\mu$m and 20 dissipation rates per decade, which raises it by up to 0.04 over the grid of table~S4. On this grid the combined estimator is within 25\% of the true value for every distribution with $D_m>1.52D_{1/2}$ (plates) and $1.44D_{1/2}$ (columns), with either closure, and there $\Kdp$ alone is within 27\%--28\%. $\Kdp$ alone is within 25\% for $D_m>1.56D_{1/2}$ and $1.49D_{1/2}$. These ratios are sufficient, not necessary. On the same grid, for $1\le D_m/D_{1/2}<1.5$ the combined estimator is within 25\% in 43\% of the cases for plates and 51\% for columns (41\% and 51\% with the superposition), among them the plates of table~3 with $D_m=40~\mu$m at $\eps=10^{-4}$~m$^2$~s$^{-3}$ ($D_m/D_{1/2}=1.32$, 0.77). With uniform dissipation the smallest ratios at which it reaches that accuracy are 0.98 (plates) and 0.85 (columns). For $0.8<D_m/D_{1/2}<1.2$ the combined estimator raises the $\Kdp$ value by factors of 1.1--2.5 and returns 0.38--0.88 of the ice water content. Table~S5 lists the smallest ratio among all cases whose $\Zdr$ exceeds a threshold. For plates the ratio can be about one half even close to the broadside $\Zdr$, for broad distributions of small crystals in weak turbulence, where Brownian rotation randomizes the smallest crystals. For columns it is at least 0.74 when $\Zdr\ge2$~dB.

The relative error of $1-Z_{\rm dr}^{-1}$ caused by an error $\sigma$ (dB) in $\Zdr$ is $(\ln10/10)\,\sigma/(Z_{\rm dr}-1)$, with $Z_{\rm dr}$ the linear ratio. For $\sigma=0.1$~dB it is 49\%, 19\%, 9\%, 4\% and 1.5\% at $\Zdr=0.2$, 0.5, 1, 2 and 4.1~dB. It reaches 25\% at 0.38~dB for $\sigma=0.1$~dB and at 0.73~dB for $\sigma=0.2$~dB. Single plates have $S=0.06$, 0.11, 0.14, 0.27 and 0.52 at $\Zdr=0.2$, 0.38, 0.5, 1 and 2~dB, with $\Gamma=0.76$, 0.77, 0.78, 0.81 and 0.87. Single columns have $S=0.10$, 0.19, 0.25, 0.49 and 0.95 and $\Gamma=0.86$, 0.87, 0.88, 0.91 and 0.99 at the same $\Zdr$.

\begin{table*}[tp]
\caption{As table~3 (entries: $\Kdp$ / combined), for $\mu_D=0$ and 6 with uniform dissipation and for $\mu_D=2$ with the lognormal superposition.}
\begin{center}
\small
\begin{tabular}{lccccccc}
\hline\hline
 & \multicolumn{3}{c}{plates, $\eps$ (m$^2$~s$^{-3}$)} & & \multicolumn{3}{c}{columns, $\eps$ (m$^2$~s$^{-3}$)}\\
 & $10^{-6}$ & $10^{-4}$ & $10^{-2}$ & & $10^{-6}$ & $10^{-4}$ & $10^{-2}$\\
\hline
\multicolumn{8}{l}{uniform dissipation (W), $\mu_D=0$}\\
$D_m=20~\mu$m & 0.43 / 0.48 & 0.18 / 0.28 & 0.02 / 0.14 & & 0.26 / 0.36 & 0.08 / 0.22 & 0.01 / 0.17\\
$D_m=30~\mu$m & 0.73 / 0.74 & 0.46 / 0.52 & 0.10 / 0.27 & & 0.56 / 0.61 & 0.27 / 0.41 & 0.04 / 0.29\\
$D_m=40~\mu$m & 0.89 / 0.90 & 0.71 / 0.75 & 0.25 / 0.45 & & 0.79 / 0.82 & 0.50 / 0.62 & 0.09 / 0.46\\
$D_m=50~\mu$m & 0.98 / 0.98 & 0.90 / 0.92 & 0.45 / 0.67 & & 0.94 / 0.95 & 0.74 / 0.83 & 0.16 / 0.64\\
\multicolumn{8}{l}{uniform dissipation (W), $\mu_D=6$}\\
$D_m=20~\mu$m & 0.46 / 0.57 & 0.13 / 0.34 & 0.01 / 0.28 & & 0.22 / 0.43 & 0.05 / 0.34 & 0.01 / 0.33\\
$D_m=30~\mu$m & 0.83 / 0.85 & 0.47 / 0.59 & 0.07 / 0.31 & & 0.61 / 0.70 & 0.23 / 0.46 & 0.03 / 0.36\\
$D_m=40~\mu$m & 0.96 / 0.96 & 0.76 / 0.81 & 0.21 / 0.46 & & 0.85 / 0.88 & 0.49 / 0.66 & 0.08 / 0.48\\
$D_m=50~\mu$m & 0.99 / 0.99 & 0.93 / 0.94 & 0.43 / 0.67 & & 0.96 / 0.97 & 0.75 / 0.85 & 0.16 / 0.66\\
\multicolumn{8}{l}{lognormal superposition (MA), $\mu_D=2$}\\
$D_m=20~\mu$m & 0.46 / 0.53 & 0.26 / 0.40 & 0.08 / 0.28 & & 0.26 / 0.40 & 0.13 / 0.33 & 0.04 / 0.27\\
$D_m=30~\mu$m & 0.78 / 0.80 & 0.57 / 0.64 & 0.24 / 0.44 & & 0.60 / 0.66 & 0.37 / 0.53 & 0.13 / 0.41\\
$D_m=40~\mu$m & 0.93 / 0.93 & 0.79 / 0.83 & 0.43 / 0.62 & & 0.82 / 0.85 & 0.61 / 0.72 & 0.25 / 0.58\\
$D_m=50~\mu$m & 0.99 / 0.99 & 0.93 / 0.95 & 0.63 / 0.80 & & 0.95 / 0.96 & 0.81 / 0.88 & 0.40 / 0.75\\
\hline\hline
\end{tabular}
\end{center}
\end{table*}

\begin{table*}[tp]
\caption{Range of the ice water content retrieved by the combined estimator and by $\Kdp$ alone (fractions of the true value) over all gamma distributions ($\mu_D=0$--6, $D_m=15$--55~$\mu$m in steps of 5~$\mu$m) and dissipation rates, sorted by $D_m/D_{1/2}$; uniform dissipation. Dissipation rates for which $D_{1/2}$ exceeds 65~$\mu$m (columns at $\eps\ge5\times10^{-3}$, plates at $\eps\ge5\times10^{-2}$~m$^2$~s$^{-3}$) are excluded.}
\begin{center}
\begin{tabular}{lccccc}
\hline\hline
 & \multicolumn{2}{c}{plates} & & \multicolumn{2}{c}{columns}\\
$D_m/D_{1/2}$ & combined & $\Kdp$ & & combined & $\Kdp$\\
\hline
0.0--0.6 & 0.120--0.355 & 0.002--0.130 & & 0.148--0.426 & 0.003--0.143\\
0.6--0.8 & 0.234--0.596 & 0.085--0.298 & & 0.236--0.672 & 0.076--0.307\\
0.8--1.0 & 0.376--0.749 & 0.245--0.524 & & 0.375--0.802 & 0.254--0.516\\
1.0--1.2 & 0.490--0.864 & 0.437--0.759 & & 0.513--0.878 & 0.448--0.734\\
1.2--1.4 & 0.618--0.924 & 0.585--0.877 & & 0.636--0.925 & 0.592--0.855\\
1.4--1.7 & 0.721--0.970 & 0.700--0.957 & & 0.743--0.970 & 0.714--0.952\\
1.7--2.2 & 0.830--0.993 & 0.819--0.991 & & 0.888--0.994 & 0.872--0.991\\
$\ge2.2$ & 0.946--0.999 & 0.943--0.999 & & 0.988--0.995 & 0.985--0.993\\
\hline\hline
\end{tabular}
\end{center}
\end{table*}

\begin{table*}[tp]
\caption{Smallest ice water content retrieved by the combined estimator among all gamma distributions and dissipation rates with $\Zdr$ (horizontal incidence) at or above a threshold, with the $\Kdp$ value for the same case and the case parameters; uniform dissipation. Broadside $\Zdr$ is 4.13~dB for plates and 2.11~dB for columns.}
\begin{center}
\begin{tabular}{lrrrrrr}
\hline\hline
shape & $\Zdr\ge$ (dB) & combined & $\Kdp$ & $\mu_D$ & $D_m$ ($\mu$m) & $\eps$ (m$^2$~s$^{-3}$)\\
\hline
plates & 0.5 & 0.13 & 0.02 & 0 & 15 & $7.9\times10^{-4}$\\
plates & 1.0 & 0.17 & 0.06 & 0 & 15 & $1.3\times10^{-4}$\\
plates & 1.5 & 0.21 & 0.10 & 0 & 15 & $3.2\times10^{-5}$\\
plates & 2.0 & 0.26 & 0.16 & 0 & 15 & $7.9\times10^{-6}$\\
plates & 2.5 & 0.31 & 0.22 & 0 & 15 & $1.3\times10^{-6}$\\
plates & 3.0 & 0.40 & 0.33 & 0 & 20 & $1.0\times10^{-5}$\\
plates & 3.5 & 0.51 & 0.46 & 0 & 20 & $3.2\times10^{-7}$\\
columns & 0.5 & 0.18 & 0.06 & 0 & 15 & $1.6\times10^{-5}$\\
columns & 1.0 & 0.24 & 0.13 & 0 & 15 & $1.0\times10^{-7}$\\
columns & 1.5 & 0.38 & 0.29 & 0 & 20 & $2.5\times10^{-7}$\\
columns & 2.0 & 0.74 & 0.71 & 0 & 35 & $4.0\times10^{-7}$\\
\hline\hline
\end{tabular}
\end{center}
\end{table*}

\section{Observed tilt widths and aerodynamic tilt models}

Glints and specular lidar returns constrain the tilt of the plates that produce them. \citet{breon2004horizontally} fitted POLDER glints with a Gaussian density of plate normals $\propto\exp(-\theta^2/\Theta^2)$, for which $\sigma_{\rm eq}=\Theta/\sqrt2$. They found $\Theta$ close to $1^\circ$ in most clouds and values below $0.5^\circ$ virtually absent above 300~hPa. Values above $3^\circ$ occurred almost only in low clouds, at pressures above 700~hPa, where turbulence is typically strong. Scanning polarization lidars give maximum tilt angles of about $1^\circ$ in clouds colder than $-30^\circ$C and $2^\circ$ in midlevel clouds warmer than $-20^\circ$C \citep{noel2005study}. Their backscatter decays exponentially with the tilt of the beam, with e-folding angles $w$ of $0.7^\circ$--$2.7^\circ$ \citep{kokhanenko2020scanning}. For an exponential density of normals $\langle\theta^2\rangle=6w^2$, so that $\sigma_{\rm eq}=\sqrt3\,w=1.2^\circ$--$4.6^\circ$. Glints observed from deep space imply tilts of about a degree, again with an exponential angular profile \citep{kostinski2026constraining}, and \citet{okamoto2020wavelength} take $0.5^\circ$--$1^\circ$ as typical, following \citet{sassen2001midlatitude}.

In the overdamped theory $\sigma_{\rm eq}=1^\circ$ requires plates of 48, 57, 74, 98 and 131~$\mu$m at $\eps=10^{-6}$, $10^{-5}$, $10^{-4}$, $10^{-3}$ and $10^{-2}$~m$^2$~s$^{-3}$. $\sigma_{\rm eq}=2^\circ$ requires 40--109~$\mu$m and $\sigma_{\rm eq}=0.5^\circ$ requires 58--156~$\mu$m over the same range. Thinner plates fall more slowly at a given maximum dimension and need larger sizes, 127 and 176~$\mu$m at $\eps=10^{-4}$~m$^2$~s$^{-3}$ for $\kappa=0.1$ and 0.05. With the thickness relation $h/d=2.01\,d^{-0.551}$ ($d$ in micrometers) adopted by \citet{breon2004horizontally}, $\Theta=1^\circ$ requires plates of 63, 81 and 117~$\mu$m at $\eps=10^{-6}$, $10^{-5}$ and $10^{-4}$~m$^2$~s$^{-3}$, whose $(a_\perp a_\parallel)^{1/2}$ of 14--22~$\mu$m lies inside or just beyond the edge of the overdamped window. Tilt widths of about a degree therefore belong to plates of about 50--120~$\mu$m in typical upper-tropospheric turbulence, and to larger ones if the plates are thin. These sizes lie in the upper part of the window or beyond it. They agree with the lower limit of tens of micrometers that \citet{kostinski2026constraining} placed on glinting plates and with the sizes above 100--200~$\mu$m that \citet{sassen1980remote} found necessary for light pillars. Because particle inertia broadens the tilt of crystals beyond about 90~$\mu$m (section~S1; \citealt{pumir2026statistical}), the larger of these sizes are lower estimates, whereas between 65 and about 87~$\mu$m inertia slightly narrows the tilt. Superposing Gaussian tilt densities, over intermittent dissipation or over a distribution of sizes with size-dependent widths, sharpens the peak of the angular profile and lengthens its tail relative to a single Gaussian, in the direction of the exponential profiles that are observed.

\textit{Size-dependent habit.} The main text uses one aspect ratio, $\kappa=1/\sqrt7$, for all plates. Natural plates become thinner with size. With the relation $h/\Dmax=2.02\,\Dmax^{-0.551}$ ($\Dmax$ in micrometers) of \citet{auer1970dimension}, $\kappa=0.39$, 0.31, 0.27 and 0.21 at $\Dmax=20$, 30, 40 and 60~$\mu$m. Thinner plates fall more slowly and align less. At $\eps=10^{-4}$~m$^2$~s$^{-3}$ a 30-$\mu$m plate falls at 1.09 instead of 1.32~cm~s$^{-1}$, and $S$ drops from 0.49 to 0.35. A 40-$\mu$m plate falls at 1.67 instead of 2.34~cm~s$^{-1}$, with $S=0.81$ instead of 0.92 and $\sigma_{\rm eq}=15.4^\circ$ instead of $9.7^\circ$. A 60-$\mu$m plate falls at 3.03 instead of 5.27~cm~s$^{-1}$, with $S=0.978$ instead of 0.995 and $\sigma_{\rm eq}=4.9^\circ$ instead of $2.2^\circ$. Because the larger plates are the thinner ones, the half-alignment size grows faster with $\eps$. Without Brownian rotation $D_{1/2}=16.2$, 32.6 and 68.4~$\mu$m at $\eps=10^{-6}$, $10^{-4}$ and $10^{-2}$~m$^2$~s$^{-3}$, against 16.9, 30.0 and 53.3~$\mu$m for $\kappa=1/\sqrt7$, and the exponent of equation~(13) rises from 0.125 to 0.157. With Brownian rotation the values are 20.1, 32.9 and 68.4~$\mu$m, against 20.2, 30.2 and 53.3~$\mu$m. Up to $\eps=10^{-4}$~m$^2$~s$^{-3}$ the change of $D_{1/2}$ stays below 10\%. The polarizabilities also vary with size, so equation~(26) holds only approximately for a size distribution of such plates.

The oriented plates that \citet{wu2025assessment} retrieved from radar Doppler velocities, 1.0--1.8~mm across with $Re_p=28$--88, and those of 0.4--1.2~mm, depending on habit, that \citet{westbrook2010doppler} inferred from a mean lidar Doppler fall speed of 0.23~m~s$^{-1}$ lie far outside the window. Extrapolated to them, the aligned-limit form of equation~(11) gives tilt widths of $0.1^\circ$--$0.2^\circ$ for the latter at $\eps=3\times10^{-4}$~m$^2$~s$^{-3}$, the largest rate in their case study, and hundredths of a degree for the former at $10^{-3}$~m$^2$~s$^{-3}$. \citet{westbrook2010doppler} reached the same conclusion by comparing the velocity difference across a 1-mm crystal at that dissipation rate, about 5~mm~s$^{-1}$, with the fall speed. In the present theory the comparison is between the Kolmogorov velocity, about 9~mm~s$^{-1}$ in that case, and the fall speed, because the tilt is set by the ratio of the Kolmogorov shear rate to the inertial alignment rate, of order $W_b^2/\nu$. The two criteria agree within a factor of about 2 for millimeter crystals but differ by the factor $\eta/\Dmax$, 70 or more, for the crystals of the overdamped window in upper-tropospheric turbulence. The randomization of millimeter crystals in strong turbulence lies beyond the present description and may involve particle inertia and unsteady fall. Unsteady fall sets in near $Re_p\approx47$ for natural hexagonal plates \citep[as reported by \citealt{wu2025assessment}]{kajikawa1992observations} and at $Re_p\approx237$ for solid hexagonal analogues in tank experiments, in which spiralling analogues incline by $7^\circ$--$28^\circ$, depending on shape \citep{stout2024stable}.

Earlier estimates of the tilt of plates in turbulence give a turbulent mean-square tilt proportional to $d^2\eps/(\nu u^2)$, the squared ratio of the velocity difference across a plate of diameter $d$ to its fall speed $u$ \citep{klett1995orientation,breon2004horizontally}. \citet{klett1995orientation} used the small-$Re$ orienting torque of nearly spherical particles \citep[as noted by][]{katz1998subsuns}. In the model of \citet{breon2004horizontally} a restoring torque $\beta\sin2\theta$, with $\beta=2C_0\rho_fd^3u^2$ and $C_0\approx0.2$ from experiments, balances a white stochastic torque. Its strength follows from the fluctuation--dissipation theorem with the turbulent kinetic energy at the particle scale, $\rho_fd^3u_t^2/2$ with $u_t=d(\eps/\nu)^{1/2}$ for crystals smaller than the Kolmogorov length \citep{klett1995orientation}, added to the thermal energy. The turbulent part of the mean-square tilt is then $u_t^2/(4C_0u^2)$. For sub-Kolmogorov crystals the turbulent rotation is instead kinematic: the velocity gradient turns the crystal at a rate of order $\tau_K^{-1}$ whatever its inertia, and in the aligned limit equation~(7) gives the mean-square tilt $2\sigma_{\rm eq}^2=(5+3\mathcal{B}^2)\nu\eps/(30F_b^2W_b^4)$. The ratio of the two is $8C_0(5+3\mathcal{B}^2)Re_p^{-2}/(60F_b^2)\approx5.8\,Re_p^{-2}$ for the plates considered here. Written in the form of \citet{breon2004horizontally}, the small-$Re_p$ inertial torque \citep{dabade2015effects} has $C_0=0.07$ for these plates and 0.11 for thin plates, and with matched torques the ratio becomes $2.0\,Re_p^{-2}$ and about $Re_p^{-2}$, respectively. The coefficients that \citet{katz1998subsuns} measured for disks sinking in oil, at Reynolds numbers of 0.1--0.25 based on the radius, correspond to $C_0=0.08$--0.13 in this form, close to the small-$Re_p$ values. The estimates coincide near $Re_p\approx1$--2.4, beyond the range of validity of the small-$Re_p$ torque, but inside the overdamped window, where $Re_p\lesssim0.1$, the equipartition model underestimates the turbulent tilt variance by more than two orders of magnitude. With it, \citet{breon2004horizontally} reproduced glint tilts of about $1^\circ$ for $\eps\approx10^{-2}$~m$^2$~s$^{-3}$, noting that such values are measured at scales larger than the crystals. Because the tilt scales as $\eps^{1/2}$, the same model gives about a tenth of that at $10^{-4}$~m$^2$~s$^{-3}$. The overdamped closure gives tilts of about a degree at the dissipation rates typical of the upper troposphere, for plates of about 50--120~$\mu$m (above). The equipartition model also places the onset of alignment at $Re_p\approx0.4$, below which the plate is assumed to keep its initial orientation, whereas the inertial torque aligns plates of $\Dmax=20$--60~$\mu$m, with $Re_p=0.003$--0.09, at rates of 0.2--14~s$^{-1}$ (table~1). In the overdamped regime the size threshold for alignment is therefore set by the turbulence, through $D_{1/2}(\eps)$, rather than by a critical Reynolds number.

The laboratory evidence at these Reynolds numbers is mixed. Thin disks sinking in oil turned horizontal at rates close to those of the small-$Re$ torque \citep{katz1998subsuns}. Hexagonal ice plates of 50--250~$\mu$m falling at Reynolds numbers of 0.1--1 through a thermally stratified tube, however, were reported to fall in random orientation \citep{burgesser2018laboratory}, although with their measured fall speeds the inertial torque would align them within a few millimeters of fall. The cause is not known. Asymmetries that separate the centers of mass and drag, which \citet{katz1998subsuns} suggested slowed or prevented the alignment of some of his disks, could act in the same way on natural crystals. If the observation holds for atmospheric crystals, a randomizing effect absent from the closure acts at $Re_p<1$, and the alignment computed here is an upper estimate.

\textit{Mass asymmetry.} Suppose the center of mass of a plate lies a distance $e$ from its center of drag, in the plane of the plate. Gravity then exerts the torque $(\rho_p-\rho_f)Vge\cos\varphi$ about the center of drag, against the inertial torque $F_D\rho_fW_b^2a_{\max}^3\sin\varphi\cos\varphi$ of appendix~A. The plate settles at the tilt $\sin\varphi=(\rho_p-\rho_f)Vge/(F_D\rho_fW_b^2a_{\max}^3)$. At fixed shape $W_b\propto\Dmax^2$ and $V\propto\Dmax^3$, so the offset that gives a fixed tilt grows as $\Dmax^4$, or as $\Dmax^3$ relative to the diameter. At 10~km a tilt of $10^\circ$ requires $e=1.6$, 5.2 and 26~nm for plates of 30, 40 and 60~$\mu$m, that is $5\times10^{-5}$, $1.3\times10^{-4}$ and $4\times10^{-4}$ of the diameter, and 0.20~$\mu$m ($2\times10^{-3}$) for a 100-$\mu$m plate. At sea level a 40-$\mu$m plate needs 10~nm and a 100-$\mu$m plate 0.40~$\mu$m. If the thickness of a disk-shaped plate varies linearly across it, with the two edges differing by 1\%, the center of mass moves by $\Dmax/1600$. For a 40-$\mu$m plate this is 25~nm, five times the offset that tilts it by $10^\circ$. Plates near the half-alignment size therefore align only if their mass is distributed symmetrically to a small fraction of a percent. Asymmetry is one candidate for the random orientation observed by \citet{burgesser2018laboratory}.

Brownian rotation sets the other limit. \citet{katz1998subsuns} estimated an rms Brownian tilt of about 0.01~rad for a plate of 30-$\mu$m radius and 3-$\mu$m thickness, taking the coefficient $C_o$ of his aligning torque to be one. His measured values, $C_o=0.17$--0.26, raise this estimate to about $1.5^\circ$--$2^\circ$. For a spheroid of the same diameter, volume and fall speed at sea level the closure gives $1.3^\circ$ per tilt component, or $1.8^\circ$ with his definition of the dispersion. In the closure Brownian rotation alone keeps $\sigma_{\rm eq}$ below 0.01~rad only for $\Dmax\ge53~\mu$m ($\kappa=1/\sqrt7$), 74~$\mu$m ($\kappa=0.1$) and 89~$\mu$m ($\kappa=0.05$) at 10~km, and turbulence tilts plates of these sizes more than Brownian rotation does once $\eps$ exceeds $2.5\times10^{-6}$, $4\times10^{-7}$ and $1.4\times10^{-7}$~m$^2$~s$^{-3}$. The lower size limit of well-aligned plates is thus set jointly by Brownian rotation and turbulence, at several tens of micrometers, consistent with the limit inferred from glints \citep{kostinski2026constraining}.

\section{Radiative and retrieval implications}

Polarized microwave extinction shows how the alignment could be constrained: the ratio of 1.4 between horizontally and vertically polarized optical thickness found by \citet{barlakas2021introducing} for the 166-GHz channel of the Global Precipitation Measurement Microwave Imager corresponds to a dichroism of $-33\%$. At the $49^\circ$ incidence angle of that channel, the Rayleigh dichroism of broadside $|\mathcal{B}|=0.75$ plates is $-43\%$, implying $S\approx0.78$, whereas broadside columns of the same aspect ratio reach only $-26\%$. This comparison is illustrative, because the observed signal integrates over habits, sizes, and radiative transfer. Lidar cloud-phase algorithms, scattering databases, and radiative-transfer models can replace a fixed tilt width or a prescribed fraction of oriented plates with the size- and turbulence-dependent relation $\sigma_{\rm eq}(\Dmax,\eps)$. The database of \citet{saito2019oriented}, which begins at $\Dmax=50~\mu$m, averages single-scattering properties over a prescribed tilt distribution; consequently, simulated lidar signals depend on both that distribution and the assumed oriented fraction. Similarly, \citet{zhou2012study} reproduced CALIPSO returns using oriented fractions of 0\%--6\% in thick mixed-phase clouds and 0\%--3\% in ice clouds warmer than $-35^\circ$C. Radiative-transfer calculations by \citet{klotzsche2006influence} assigned the same maximum tilt angle, with a uniform distribution below it, to all crystal sizes. They found that treating horizontally oriented columns as randomly oriented overestimates the reflected solar flux at high Sun and underestimates it by as much as 104~W~m$^{-2}$ at intermediate solar zenith angles. For small plates the reflected flux can be overestimated by up to 155~W~m$^{-2}$. These were regarded as upper limits because real populations are neither entirely horizontal nor entirely random, and the authors proposed linking orientation to crystal size through the surrounding turbulent flow or electric fields. Their nearly horizontal case, with a maximum tilt of $1^\circ$, corresponds to $\sigma_{\rm eq}=0.41^\circ$ for plates and $0.58^\circ$ for columns. Our closure reaches these widths only for plates of 62--93~$\mu$m and columns of 76--120~$\mu$m, mostly beyond the overdamped window, at $\eps=10^{-6}$--$10^{-4}$~m$^2$~s$^{-3}$, whereas plates of 27--51~$\mu$m and columns of 33--65~$\mu$m have widths equivalent to maximum tilts of $10^\circ$--$20^\circ$. Because the small-crystal distribution considered by \citet{klotzsche2006influence}, with an effective radius of 25~$\mu$m, spans these sizes, no single maximum tilt can represent its orientation state. \citet{breon2004horizontally}, in contrast, inferred from glints, which preferentially sample the upper cloud, an oriented fraction of only 0.1\%--1\% when weighted by crystal area. Because orientation changes the sunlight intercepted by a plate by at most a factor of 2, they concluded that this fraction would have little effect on cloud albedo. At Summit, Greenland, however, \citet{stillwell2019radiative} found a shortwave cloud radiative effect at the surface about 22\% larger at a given solar zenith angle when oriented crystals were present. The radiative impact therefore need not always be negligible, although the comparison does not isolate orientation from other cloud differences. Glints sample only plates tilted by roughly a degree or less and therefore, within the overdamped model, preferentially detect the largest crystals. Crystals near the alignment transition instead have tilt widths of several degrees and can contribute to specular lidar returns at off-nadir angles without producing strong glints. The two observationally based estimates therefore bracket, rather than describe, the population emphasized here: small crystals whose orientation width is controlled by the local dissipation rate.

Finite size adds a radiative consequence whose sign depends on geometry. To second order in size parameter the Rayleigh--Gans cross sections and asymmetry parameter depend on orientation through $\PL{2}P_2(\cos\theta_i)$, so crystals aligned by weak turbulence scatter less strongly forward than random ones for radiation incident within $54.7^\circ$ of the vertical and more strongly beyond it. The T-matrix shows the same for solid ice at $\Xm=1$, with the crossing near $53^\circ$ (figure~5a of the main text). For solid ice up to $\Xm=1$, the asymmetry parameter of aligned plates at vertical incidence is a quarter of the random-orientation value. An orientation assumption fixed independently of turbulence therefore biases the asymmetry parameter in opposite directions for near-nadir and limb paths at wavelengths below about 0.4~mm, where the crystals of the overdamped window reach $\Xm$ of order one. For isotropic radiation the $O(X^2)$ effect averages out, since the mean of $P_2(\cos\theta_i)$ over the sphere vanishes, but it survives for directional beams and in hemispheric fluxes, where the cosine weighting leaves a quarter of its vertical-incidence value. The $53^\circ$ zenith angle of conical-scanning radiometers lies close to the angle at which these effects vanish, which is why the polarization-averaged extinction there hardly depends on orientation (section~3a of the main text) and the polarization difference carries the orientation signal. This angular structure belongs to size parameters of order one. At solar wavelengths the same crystals are hundreds of wavelengths across, and ray paths rather than interference set the response. Horizontally oriented columns then scatter more strongly forward than random ones at high Sun, the opposite of the result at $\Xm=1$. They scatter least strongly forward near a solar zenith angle of $50^\circ$, where retroreflection and internal reflection dominate \citep{klotzsche2006influence}.

\section{Numerical methods and verification}

Moments of the Boltzmann density were evaluated in closed form with the error and Dawson functions, with a series below $b=0.05$, and independently by Gauss--Legendre quadrature in the angle from the attractor on panels graded with width $b^{-1/2}$. The two agree to $10^{-10}$ in relative terms for $10^{-6}\le b\le10^5$ and reproduce the random limits $\langle\varphi^2\rangle=\pi^2/4-2$ and $\pi-2$. The lognormal superposition (8) was evaluated with 120-point Gauss--Hermite quadrature in $\ln\phi$. Doubling the node count changes the moments by less than $10^{-6}$, and the aligned-limit moments reproduce $\langle\mu^2\rangle=1/2b$ and the flatness $3e^{\slnphi}$ of \citet{roy2026orientation}. The Rayleigh averages (B1) agree with direct quadrature over $(\mu,\zeta)$ to better than $10^{-13}$. They agree with T-matrix calculations at a volume-equivalent size parameter of 0.02 ($\Xm=0.028$ for plates and 0.038 for columns) to within 0.0011~dB in $\Zdr$ and 0.005~dB in LDR, the residual being the expected $O(\Xm^2)$ dynamic correction.

The T-matrix calculations used the extended-boundary-condition code of \citet{mishchenko1998capabilities}, compiled through the interface of \citet{leinonen2014high}, with a convergence tolerance of $10^{-4}$. For a sphere with $m=1.78+0.005i$ at size parameter one the scattering and extinction cross sections agree with Mie theory to seven digits. For real $m$ the angular integral of the phase matrix reproduces the optical-theorem extinction to $10^{-5}$. As $m\to1$ the Rayleigh-normalized backscatter approaches the Rayleigh--Gans average (20) with differences of 2.2\%, 0.43\% and 0.09\% at $m-1=0.05$, 0.01 and 0.002. The Legendre-weight average (22) agrees with direct quadrature on 400 graded nodes to better than $10^{-13}$ for $b=0.5$, 5 and 500. A $(2N+1)$-point Gauss--Legendre rule on $[0,1]$, by contrast, integrates only to degree $4N+1$ and is not exact. The azimuthal averages used $2N+6$ (vertical incidence) and $4N+8$ (oblique incidence) equally spaced crystal azimuths, which integrate the trigonometric kernels exactly, and the angular integrals used $N+12$ Gauss--Legendre nodes in $\cos\vartheta$ and $2N+8$ azimuths. The truncation orders were $N=5$--7 for both shapes at $\Xm\le2$. Replacing $m=1.79+0.01i$ by the indices of \citet{warren2008optical} at 0.2 and 0.1~mm, $1.81+0.05i$ and $1.87+0.17i$, changes the departures $|f_Q-S|$ at $\Xm=1$, including those of $g$ and the LDR, by at most 0.004 and the broadside-to-random backscatter ratio by less than 0.1~dB. At $\Xm=2$ the added absorption reduces the departure of the plate extinction from 0.14 to 0.05.

The Rayleigh--Gans averages (20) used coefficients $R_n$ up to $n=80$ from 256-point Gauss--Legendre quadrature in $\mu_q$, 192 Gauss--Legendre nodes in $\cos\vartheta$ and 96 azimuths $\xi$. For Boltzmann densities with $(b,\theta_i)=(2,0^\circ)$, $(8,40^\circ)$ and $(8,90^\circ)$ they agree with direct quadrature over the crystal orientation to six digits in $\sigma_s/\sigma_0$ and $g$. At $\Xm=0.1$ the exact values of $1-\sigma_s/\sigma_0$, $g$ and $1-\sigma_b/\sigma_{b0}$ differ from (21) by less than 0.4\%, as expected for errors of $O(X^4)$, and at $\theta_i=54.7^\circ$ they vary by less than 0.03\% between $b=0.5$, 5 and 50. The T-matrix asymmetry parameter at oblique incidence was integrated over 40 Gauss--Legendre nodes in the polar angle of the scattered direction and 72 azimuths, with 16 crystal azimuths for the columns. At vertical incidence it reproduces the values from the Legendre-weight average to five digits. The angles at which broadside and random orientation give the same value were found by root finding. For $m=1.01+10^{-4}i$ they agree with the exact Rayleigh--Gans crossings to $0.03^\circ$ at $\Xm=0.25$ and 1, and the broadside-to-random ratios of $g$ to 0.002. For ice the asymmetry-parameter crossing lies at $53.80^\circ$, $53.75^\circ$, $53.58^\circ$ and $52.95^\circ$ for plates and $53.57^\circ$, $53.52^\circ$, $53.37^\circ$ and $52.79^\circ$ for columns at $\Xm=0.1$, 0.25, 0.5 and 1. The scattering cross section crosses at $54.73^\circ$ for both shapes at $\Xm=0.1$, as the Rayleigh limit requires, and at $53.9^\circ$ (plates) and $54.1^\circ$ (columns) at $\Xm=1$, whereas the backscatter, which contains $\PL{4}$ already in the Rayleigh limit, crosses at $53.4^\circ$ and $53.9^\circ$ at $\Xm=0.1$. A sphere with $m=1.79+0.01i$ has $\sigma_b/\sigma_b^{\rm R}=1-0.25x^2$ for $x\le0.2$, and one with $m=1.01$ has $1-0.79x^2$.

The specular enhancement (27) uses the plate density normalized on $0\le\mu\le1$, whose integral is one to $10^{-12}$ for $10^{-3}\le b\le5\times10^3$ and which reduces to the Gaussian form for large $b$. The lognormal superposition uses 200 Gauss--Hermite nodes, and halving their number changes $G$ by less than $10^{-9}$ in relative terms. The diffraction lobe is applied by quadrature over the angle from the beam, with the azimuthal average of $\psi$ computed on 256 azimuths for $b\le300$ and from the Bessel average of the Gaussian form for larger $b$. For 60-$\mu$m plates with $\sigma_{\rm eq}=0.45^\circ$ at 532~nm the result agrees with a direct two-dimensional convolution with the Airy pattern to 0.2\% at $0.3^\circ$, $1^\circ$ and $1.5^\circ$ off nadir.

These checks verify the numerical evaluation of the model. The accuracy of the closure itself, equation~(7), is that established against stochastic simulations of the same forcing model by \citet{roy2026orientation}, with a largest error of 15\% on $\langle\mu^2\rangle$ near $\Lambda\approx0.45$ for a slender column (16\% with the full velocity-gradient forcing). Where the torque is weak the closure underestimates the alignment (section~2a).

\section{EarthCARE data, selection and sensitivity}

The frames were taken on the 1st and 15th of each month from December 2024 to November 2025. Of these 24 days, 22 had matching products; 15 December 2024 and 1 April 2025 did not. We used up to two orbits a day, about 12~h apart, and the frames of each orbit for which all three products were available, by day and night and at all latitudes. Of the 276 selected frames, 273 could be processed. We used the lidar ratio, the particle linear depolarization ratio, the extinction and their errors from the ESA A-EBD product \citep[processing baselines BA and BB;][]{donovan2024earthcare,esa2025atlebd} at its medium resolution of 11 along-track pixels of the joint grid, together with the synergetic radar--lidar target classification \citep{irbah2023classification,esa2025actc}. From the JAXA AC-CLP product at 10~km \citep{sato2025jaxa,jaxa2025acclp} we used the same cloud quantities, the temperature, the ATLID cloud particle type, the effective radius, the radar reflectivity and the radar--lidar detection flag. The effective radius is a melted-equivalent radius weighted towards large crystals, retrieved from radar and lidar or from the radar alone. A-EBD bins were kept with quality status good or likely good, errors of $S_{\rm lid}$ and $\delta$ below 5~sr and 0.05, and a synergetic class containing ice, snow or supercooled liquid; AC-CLP bins with an ATLID cloud type. Temperatures were attached to the A-EBD bins by matching times and heights. This leaves 2.2 million A-EBD and 6.2 million AC-CLP cloud bins between $-50$ and $0^\circ$C, of which 0.94 and 3.2 million are low-extinction.

The ATLID target classification (A-TC) of the ESA products labels classes 102 and 103 together as ``unknown: aerosol classification outside of param space'' in the product definition of ATL\_TC\_2A; the copy of the classification in AC\_\_TC\_\_2B lists 102 but not 103. The JAXA AC-CLP variable \texttt{ATLquality\_flag\_10km}, described in the product as the ATLID quality flag method, carries values 0--7 that are not documented further in the release notes we could find.

The A-EBD lidar ratios pile up near 10~sr. At $-20$ to $-10^\circ$C, bins between 9.75 and 10.25~sr are nearly twice as frequent per 0.25~sr as those between 11 and 13~sr, and the frequency falls steeply below 9.75~sr. The specular lidar ratios of about 4~sr implied by the depolarization of the A-EBD signature bins therefore cannot be represented. Only 44\% of the A-EBD signature bins have collocated AC-CLP values of $S_{\rm lid}$ and $\delta$, and 9\% of those meet the AC-CLP criterion; their median AC-CLP values are 16~sr and 0.07. At most of the others AC-CLP detects the cloud with the radar alone. The fraction of low-extinction, weakly depolarizing ($\delta<0.06$) bins that carries the signature rises from 5\% or less below $-35^\circ$C to 17\%--19\% at $-15$ to $-10^\circ$C in A-EBD, but is about 20\%--25\% between $-45$ and $-8^\circ$C in AC-CLP, where such bins are rare below $-35^\circ$C. At $-20$ to $-10^\circ$C the ESA lidar-only classification assigns 56\% of the A-EBD signature bins to aerosol types, 16\% to supercooled water, 9\% to ice and 17\% to class 103, unknown aerosol outside the parameter space of the classification. The JAXA classification assigns 84\% of the AC-CLP signature bins to types containing horizontally oriented plates, identified from depolarization, attenuated backscatter and temperature \citep{sato2025jaxa}, and 16\% to supercooled water. The collocated AC-CLP effective radius has a median of 81~$\mu$m at the A-EBD signature bins against 92~$\mu$m at the other low-extinction bins at the same temperatures, and the reflectivity $-8.8$ against $-6.9$~dBZ. The differences persist both where the AC-CLP product detects the cloud with radar and lidar and where it detects it with the radar alone. Uncertainties of the occurrence are 5\%--95\% ranges of a bootstrap over frames.

\textit{Sensitivity tests.} Table~S6 repeats the occurrence statistics of figure~10b for variants of the selection: depolarization limits of 0.04 and 0.08 instead of 0.06; lidar-ratio limits of 12 and 15~sr (A-EBD) and 6 and 10~sr (AC-CLP); extinction limits of 0.25 and 1~km$^{-1}$; the depolarization criterion alone; the A-EBD bins at the lower limit of the retrieved lidar ratio excluded, or only lidar ratios below that limit; tighter A-EBD quality and error limits; each AC-CLP method flag separately; night and day; and the frames of each A-EBD processing baseline. In every variant that keeps a lidar-ratio criterion, apart from the subset of AC-CLP bins with method flag 2, the maximum lies in the $2.5^\circ$C bins centered at $-11.25$ or $-13.75^\circ$C. The occurrence at $-20$ to $-10^\circ$C then exceeds that at $-35$ to $-20^\circ$C by factors of 3.3--7.8 in both products. The level of the occurrence changes by up to an order of magnitude with the thresholds. Without a lidar-ratio criterion the AC-CLP occurrence peaks next to $0^\circ$C, where supercooled water dominates the weakly depolarizing bins.

Three features of the retrievals limit the interpretation. First, about half of the A-EBD signature bins at $-20$ to $-10^\circ$C (51\%) lie at the lower limit of 9.75--10.25~sr, against 7\% of the other low-extinction bins. Their lidar-ratio and depolarization errors are smaller than those of the other low-extinction bins (medians 2.0~sr and 0.019 against 3.0~sr and 0.024). Excluding the lower-limit bins halves the occurrence but leaves its temperature dependence and raises the contrast to 7.8. Second, 53\% of the AC-CLP signature bins carry method flag 2, against 2\% of all low-extinction cloud bins. Restricted to flag 1, the occurrence falls by about half and keeps its maximum near $-14^\circ$C and a contrast of 5.6. The flag-2 bins alone show low lidar ratios at all temperatures, including 0.6\% of the cloud bins at $-50$ to $-35^\circ$C. Third, no frame is available in both A-EBD baselines: BA covers December 2024 to 1 October 2025 and BB the later frames, so the two subsets also differ in season. On the same subsets the AC-CLP product, with one baseline, changes by a similar factor (A-EBD/AC-CLP peak ratio 2.2 for the BA frames and 1.9 for the BB frames). The lower limit of the A-EBD lidar ratio also persists in BB (pile-up ratio 2.2 against 1.8 in BA). We therefore find no effect of the baseline beyond the seasonal change.

The specular line of equation~(28) is the locus of a fixed ratio of extinction to cross-polar backscatter. A bin moves along it whenever copolar backscatter is added at fixed extinction and cross-polar backscatter, whether by face reflection or by an error in the copolar channel, so the position of a bin on the line does not identify specular reflection. Mixtures of randomly oriented ice and supercooled water lie between the two lidar ratios, near 18~sr for water and 24--30~sr for ice in these products, and cannot reach the signature box. The classifications cited above are derived from the same lidar ratio, depolarization and attenuated backscatter and therefore do not provide independent evidence for plates.

\begin{table*}[tp]
\caption{Sensitivity of the EarthCARE specular-signature statistics to the selection (273 frames). For each product: temperature of the maximum occurrence ($^\circ$C, center of $2.5^\circ$C bins, excluding the bin next to $0^\circ$C), maximum occurrence and occurrence at $-20$ to $-10^\circ$C (band; \% of cloud bins), and the ratio of the occurrence at $-20$ to $-10^\circ$C to that at $-35$ to $-20^\circ$C. Reference selection: extinction below 0.5~km$^{-1}$, $\delta<0.06$, $S_{\rm lid}\le10.5$~sr (A-EBD) or $S_{\rm lid}<8$~sr (AC-CLP).}
\begin{center}
\small
\setlength{\tabcolsep}{4pt}
\begin{tabular}{>{\raggedright\arraybackslash}p{12.5pc}rrrrcrrrr}
\hline\hline
 & \multicolumn{4}{c}{A-EBD} & & \multicolumn{4}{c}{AC-CLP}\\
variant & $T_{\max}$ & max & band & ratio & & $T_{\max}$ & max & band & ratio\\
\hline
reference selection & $-11.25$ & 1.64 & 1.18 & 5.2 & & $-11.25$ & 0.77 & 0.55 & 6.0\\
$\delta<0.04$ & $-11.25$ & 0.90 & 0.64 & 5.4 & & $-11.25$ & 0.43 & 0.30 & 5.4\\
$\delta<0.08$ & $-11.25$ & 2.40 & 1.81 & 5.4 & & $-13.75$ & 1.12 & 0.78 & 6.5\\
$S_{\rm lid}\le12$ (A-EBD), $<10$~sr (AC-CLP) & $-11.25$ & 3.57 & 2.50 & 6.5 & & $-11.25$ & 1.13 & 0.77 & 6.5\\
$S_{\rm lid}\le15$ (A-EBD), $<6$~sr (AC-CLP) & $-11.25$ & 5.95 & 4.23 & 6.9 & & $-11.25$ & 0.49 & 0.36 & 5.2\\
extinction $<0.25$~km$^{-1}$ & $-11.25$ & 1.00 & 0.78 & 4.5 & & $-11.25$ & 0.62 & 0.42 & 5.8\\
extinction $<1$~km$^{-1}$ & $-11.25$ & 1.95 & 1.42 & 5.1 & & $-13.75$ & 0.92 & 0.64 & 6.5\\
$\delta<0.06$ only, no $S_{\rm lid}$ criterion & $-11.25$ & 8.85 & 6.88 & 3.7 & & $-3.75$ & 3.30 & 2.25 & 5.0\\
A-EBD, $9.75\le S_{\rm lid}\le10.25$~sr excluded & $-11.25$ & 0.86 & 0.58 & 7.8 & & -- & -- & -- & --\\
A-EBD, $S_{\rm lid}<9.75$~sr & $-11.25$ & 0.52 & 0.35 & 7.6 & & -- & -- & -- & --\\
A-EBD, status 0, errors $\le2.5$~sr and 0.025 & $-11.25$ & 2.78 & 1.97 & 3.3 & & -- & -- & -- & --\\
AC-CLP, method flag 1 only & -- & -- & -- & -- & & $-13.75$ & 0.36 & 0.26 & 5.6\\
AC-CLP, method flag 2 only & -- & -- & -- & -- & & $-13.75$ & 7.23 & 6.16 & 2.0\\
night & $-11.25$ & 1.57 & 1.13 & 4.8 & & $-13.75$ & 0.61 & 0.47 & 5.9\\
day & $-11.25$ & 1.75 & 1.29 & 6.0 & & $-11.25$ & 1.03 & 0.66 & 6.1\\
frames with A-EBD baseline BA (232) & $-11.25$ & 1.74 & 1.23 & 5.3 & & $-11.25$ & 0.79 & 0.56 & 6.1\\
frames with A-EBD baseline BB (41) & $-13.75$ & 1.30 & 0.96 & 4.6 & & $-11.25$ & 0.68 & 0.48 & 5.1\\
\hline\hline
\end{tabular}
\end{center}
\end{table*}

\end{document}